\documentclass[11pt,journal,onecolumn,draftclsnofoot]{IEEEtran}

\usepackage[T1]{fontenc}
\usepackage{amsmath,amssymb,amsthm,mathtools,bm}
\usepackage{booktabs,array,longtable,graphicx,float}
\usepackage{cite}
\usepackage{xcolor,hyperref}
\usepackage[nameinlink,noabbrev]{cleveref}
\usepackage{microtype}

\hypersetup{hidelinks}
\graphicspath{{figures/}}
\numberwithin{equation}{section}

\newtheorem{assumption}{Assumption}[section]
\newtheorem{theorem}{Theorem}[section]

\newtheorem{lemma}{Lemma}[section]
\newtheorem{corollary}{Corollary}[section]
\theoremstyle{definition}
\newtheorem{remark}{Remark}[section]

\newcommand{\R}{\mathbb R}
\newcommand{\E}{\mathbb E}

\newcommand{\Cov}{\operatorname{Cov}}

\newcommand{\ABC}{\mathrm{ABC}}

\newcommand{\rev}[1]{{\color{black}#1}}
\newenvironment{revision}{\begingroup\color{black}}{\par\endgroup}

\title{Likelihood-Free Schr\"odinger--F\"ollmer Sampling
for Bayesian Inference with Fixed-Proposal Reuse}
\author{Xihan Ren\thanks{The work of Z. He was supported by the National Natural Science Foundation of China under Grant 12671500 and by the Guangdong Basic and Applied Basic Research Foundation under Grant 2025A1515011888.}\thanks{Xihan Ren: School of Mathematics, South China University of Technology,
    Guangzhou 510641, People's Republic of China.
    \texttt{202520130109@mail.scut.edu.cn}.},
  Zhijian He\thanks{Zhijian He: School of Mathematics, South China University of Technology,
    Guangzhou 510641, People's Republic of China.
    \texttt{hezhijian@scut.edu.cn}.}
 and
  Du Ouyang\thanks{Du Ouyang: Centre de Math\'{e}matiques Appliqu\'{e}es, \'{E}cole Polytechnique, Palaiseau 91128, France.\newline
     \texttt{du.ouyang@polytechnique.edu}.}
    }

\begin{document}

\maketitle

\begin{abstract}
We study Bayesian posterior sampling when the likelihood is an expectation that is costly to evaluate.
We construct a likelihood-free Schr\"odinger--F\"ollmer sampler whose drift is written as a ratio of expectations over Gaussian variables and simulator outputs and is estimated directly from joint draws of the two.
Importance sampling with a fixed proposal allows the simulator outputs to be generated once and reused across all time steps and trajectories.
For the likelihood-free Schr\"odinger--F\"ollmer sampler, which draws new Monte Carlo samples at each step, we bound the terminal Wasserstein error under regularity and moment assumptions, separating time-discretization and integration errors.
For the likelihood-free Schr\"odinger--F\"ollmer sampler with sample reuse, we prove that the output converges to the posterior in the Wasserstein distance, assuming only a finite posterior second moment.
Numerical experiments illustrate the reduction in simulator calls and the benefits of randomized quasi-Monte Carlo for drift estimation.
\end{abstract}

\section{Introduction}
\label{sec:introduction}

We consider Bayesian inverse problems, in which unknown model parameters are inferred from observed data.
The Bayesian approach combines prior information with a likelihood model and quantifies uncertainty through the posterior distribution \cite{stuart2010inverse}.
This formulation is used, for example, for parameter estimation in differential equations \cite{dashti2017bayesian} and in population genetics \cite{beaumont2002abc}.
We focus on likelihoods given by conditional expectations that are costly to evaluate, and aim to reduce the simulation cost of posterior sampling.

Let $x\in\R^d$ denote the unknown parameter.
The posterior density is
\begin{equation}
 \pi(x)=\frac{\pi_0(x)L(x)}{\mathcal Z},
 \qquad
 \mathcal Z=\int_{\R^d}\pi_0(x)L(x)\,\mathrm dx,
 \label{eq:intro_posterior}
\end{equation}
where $\pi_0$ is the prior density, $L$ is the likelihood or a specified likelihood approximation, and $0<\mathcal Z<\infty$.
We consider targets for which
\begin{equation}
 L(x)=\E_{\xi\mid x}\!\left[\psi(x,\xi)\right],
 \qquad
 \xi\sim p(\cdot\mid x),
 \qquad
 \psi(x,\xi)\ge0.
 \label{eq:intro_expectation}
\end{equation}
Here $\xi$ denotes simulator outputs or auxiliary variables, $p(\cdot\mid x)$ is their conditional distribution given $x$, and $\psi(x,\xi)$ is a computable nonnegative weight.
We assume that $\xi\sim p(\cdot\mid x)$ can be simulated, but not that the density $p(\xi\mid x)$ can be evaluated.

Two examples illustrate the representation in \eqref{eq:intro_expectation}.
In approximate Bayesian computation (ABC), simulated data receive larger weights when they agree more closely with the observations; averaging these weights gives an approximate likelihood \cite{beaumont2002abc,sisson2018handbook}.
In latent-variable models, averaging the conditional likelihood over the unobserved variables gives the exact likelihood, which can therefore be estimated without bias by simulation \cite{andrieu2009pseudomarginal}.
Section~\ref{sec:likelihood_examples} describes both constructions in detail.

Our goal is to generate approximately unweighted samples from $\pi$ when $L$ cannot be evaluated directly; such samples give any posterior summary, such as $\E_{X\sim\pi}[h(X)]$, without reweighting.
Representation \eqref{eq:intro_expectation} also permits importance-sampling estimates of posterior expectations, but these are weighted averages rather than posterior samples.
Equally weighted samples can be obtained by resampling the weighted points, but they are then confined to those points, and when the importance weights degenerate, which typically happens in high dimensions or for concentrated posteriors, only a few distinct points remain \cite{agapiou2017importance}.
On the other hand,
estimating \eqref{eq:intro_expectation} at every parameter value visited by a sampler can require many costly simulations, so we seek accurate posterior samples from fewer simulator evaluations.

Most methods for intractable likelihoods are based on Markov chain or sequential Monte Carlo.
ABC compares simulated and observed data through a kernel and is implemented by rejection sampling \cite{beaumont2002abc}, Markov chain Monte Carlo (MCMC) \cite{marjoram2003mcmc}, or sequential Monte Carlo \cite{delmoral2012abcsmc}; see Marin et~al.~\cite{marin2012abc} and Sisson et~al.~\cite{sisson2018handbook} for reviews.
Synthetic likelihood replaces the likelihood of the summaries by a Gaussian fitted to simulations \cite{price2018bsl}, and pseudo-marginal MCMC replaces the likelihood by a nonnegative unbiased estimator while still targeting the exact posterior \cite{andrieu2009pseudomarginal,andrieu2010pmcmc}.
These MCMC-based methods require new simulations at every proposed parameter and produce correlated output, and the noise in the likelihood estimate must be tuned to avoid sticky chains \cite{doucet2015efficient}.
Neural simulation-based inference instead trains a conditional density estimator on simulations \cite{cranmer2020frontier}, at the cost of network design and training.

Diffusion-based samplers transport a simple reference distribution to the target in finite time, so independent trajectories can be run in parallel without burn-in \cite{zhang2022pis,vargas2023dds}.
Most of them learn the drift with neural networks.
We build on Schr\"odinger--F\"ollmer sampling (SFS) \cite{huang2021sfs,jiao2021convergence}, a diffusion-based sampler whose drift is an explicit Gaussian expectation and can therefore be estimated by Monte Carlo without training.

SFS generates samples from a prescribed target by simulating the diffusion
\begin{equation}\label{eq:SFS}
 \mathrm dX_t=b(X_t,t)\,\mathrm dt+\mathrm dB_t,
 \qquad X_0=0,
\end{equation}
where $B_t$ is a standard $d$-dimensional Brownian motion.
Let $\varphi$ be the density of $B_1\sim\mathcal N(0,I_d)$, which serves as a Gaussian reference, set $f=\pi/\varphi$, and choose
\begin{equation}
 b(x,t)=\frac{\E_Z[f(Z)(Z-x)]}{(1-t)\E_Z[f(Z)]},
 \qquad Z\sim\mathcal N(x,(1-t)I_d),\quad 0\le t<1.
 \label{eq:intro_sfs_drift}
\end{equation}
Under suitable regularity and integrability conditions, this choice gives $X_1\sim\pi$ \cite{huang2021sfs,jiao2021convergence}; the resulting process solves the Schr\"odinger bridge problem between the point mass at the origin and $\pi$ \cite{leonard2014survey,huang2021sfs}.
Since $f(Z)=\pi_0(Z)L(Z)/(\mathcal Z\varphi(Z))$, the unknown normalizing constant $\mathcal Z$ cancels in the ratio.
However, $f$ still involves $L$, which cannot be evaluated.

We combine \eqref{eq:intro_expectation} with iterated expectation to write the drift as a ratio of joint expectations over a Gaussian variable $Z$ and an auxiliary variable $\xi\sim p(\cdot\mid Z)$.
The drift can then be estimated directly from simulated pairs $(Z,\xi)$.
Section~\ref{sec:methods} gives the derivation and the two resulting samplers.

Several works study the construction and numerical approximation of SFS.
Huang et~al.~\cite{huang2021sfs} introduced the sampler and analyzed its time discretization.
Jiao et~al.~\cite{jiao2021convergence} established quadratic Wasserstein error bounds under smoothness and boundedness assumptions on the target density ratio, separating time-discretization and Monte Carlo (MC) errors.
The MC estimator analyzed by Jiao et~al.~\cite{jiao2021convergence} writes the drift as $\E_Z[\nabla f(Z)]/\E_Z[f(Z)]$ and thus requires the gradient of the density ratio.
In \eqref{eq:intro_sfs_drift}, the derivative is instead moved onto the Gaussian kernel, so only values of $f$ are needed; this is what later allows $f$ to be replaced by simulator outputs.
Their MC error bound therefore does not cover our estimator.

Drift approximation is also related to importance sampling with random weights.
Self-normalized importance sampling estimates a ratio of integrals using a common set of samples; its consistency and asymptotic properties are discussed in Owen~\cite[Section~9.2]{owen2013mc}.
Llorente et~al.~\cite{llorente2022noisy} studied noisy importance sampling and the effect of random target evaluations on proposal design.
Riel and Bischoff~\cite{riel2025ensemble} used importance sampling and particle ensembles to approximate gradients of log densities without target gradients.
Importance sampling with mixture proposals was studied by Hesterberg~\cite{hesterberg1995} and Owen and Zhou~\cite{owen2000mixture}.

Quasi-Monte Carlo (QMC) and randomized QMC (RQMC) integration rules can improve on the MC error under suitable regularity assumptions \cite{niederreiter1992,owen1997variance,owen1997scramble}.
For the SFS drift, however, the integrands depend on the current state and time, and reusing the same samples makes the subsequent states depend on them.
An integration bound at a fixed state therefore does not by itself yield a bound on the terminal distribution.

The works above provide diffusion error estimates, importance sampling constructions, and integration rules, but they do not give a sampler for \eqref{eq:intro_expectation} that avoids estimating the likelihood at every parameter value.
Our contributions are as follows.
\begin{itemize}
 \item \textbf{Likelihood-free drift.} Using iterated expectation, we write the SFS drift as a ratio of joint expectations over Gaussian variables and simulator outputs \eqref{eq:problem_joint_drift}. The drift can thus be estimated from simulated pairs by MC or RQMC, without likelihood values or simulator derivatives.

 \item \textbf{Likelihood-free SFS.} Estimating this drift with new samples at each time step gives likelihood-free SFS. Theorem~\ref{thm:path_error} bounds its terminal quadratic Wasserstein distance $W_2$ error by $O(\sqrt{h}+\sqrt{\Lambda_h/m})$, where $h$ is the time step, $m$ is the number of samples per step, and $\Lambda_h$, defined in \eqref{eq:integrated_local_error}, accumulates the conditional mean-square drift-estimation error over the time grid.

 \item \textbf{Likelihood-free SFS with sample reuse.} Importance sampling with a fixed proposal lets one set of $M$ simulated pairs serve all time steps and trajectories, reducing the number of simulator calls from $NKm$ to $M$; in the Gaussian benchmark, $8192$ calls replace about $10^7$. Theorem~\ref{thm:reuse_posterior_terminal} proves that the output converges to the posterior in the quadratic Wasserstein distance $W_2$ under only a finite posterior second moment, and Theorem~\ref{thm:qbank_pointwise} gives a central limit theorem for the drift estimator that guides the choice of proposal.
\end{itemize}

This paper is organized as follows.
Section~\ref{sec:problem} describes the problem setting and the likelihood representations. Section~\ref{sec:methods} introduces the two samplers. Section~\ref{sec:theory} gives terminal error and convergence results for both, followed by the pointwise drift limit and the proposal-variance analysis.
Section~\ref{sec:experiments} reports numerical experiments, and Section~\ref{sec:conclusion} concludes. Appendices~\ref{sec:appendix_fresh} and~\ref{sec:appendix_reuse} contain the proofs and supporting calculations.

\section{Problem setting}
\label{sec:problem}

\label{sec:likelihood_examples}

Let $x\in\R^d$ denote the unknown parameter and let $y$ denote the observed data.
Since $y$ is fixed throughout, we suppress the dependence on $y$ in the notation for the likelihood and the posterior.
We consider the posterior density
\begin{equation}
 \pi(x)
 =
 \frac{\pi_0(x)L(x)}{\mathcal Z},
 \qquad
 \mathcal Z
 =
 \int_{\R^d}\pi_0(u)L(u)\,\mathrm du,
 \label{eq:posterior}
\end{equation}
where $\pi_0$ is a prior density and $0<\mathcal Z<\infty$.
Throughout this section, $L$ is a nonnegative function defining the target; it may be the exact likelihood or a specified approximation, such as an ABC likelihood.

We assume the representation
\begin{equation}
 L(x)
 =
 \E_{\xi\mid x}\!\left[\psi(x,\xi)\right]
 =
 \int_{\Xi}\psi(x,\xi)\,p(\mathrm d\xi\mid x),
 \qquad
 \psi(x,\xi)\ge0.
 \label{eq:likelihood_expectation}
\end{equation}
Here $\Xi$ is the space of simulator outputs or auxiliary variables,
$p(\mathrm d\xi\mid x)$ is their conditional probability law,
and $\psi(x,\xi)$ is a measurable, computable weight;
the expectation is assumed finite for all parameter values considered.
When this law has a density with respect to a reference measure,
we also write $p(\xi\mid x)$ for that density.
The samplers require draws from this law but never evaluate its density.

In ABC \cite{beaumont2002abc,blum2010abc}, the model is a simulator: for given $x$, it generates synthetic data $\xi\sim p(\cdot\mid x)$ in the same space $\Xi$ as the observed data $y$, but the likelihood $p(y\mid x)$ of the observed data cannot be evaluated.
ABC therefore compares the simulated data $\xi$ with $y$.
Let $S:\Xi\to\R^r$ be a summary map \cite{prangle2018summary}, where $r$ is the number of summary statistics. The vectors $S(\xi)$ and $S(y)$ may contain, for example, sample means or variances.
We set
\begin{equation}
 \psi(x,\xi)
 =
 K_\delta\!\left(S(\xi),S(y)\right),
 \qquad
 \xi\sim p(\cdot\mid x),
\end{equation}
where the nonnegative kernel $K_\delta$ gives larger weights to closer matches, and its bandwidth $\delta>0$ controls how much mismatch is tolerated.
Equation~\eqref{eq:likelihood_expectation} then defines the approximate likelihood $L_{\ABC}(x)=\E_{\xi\mid x}[K_\delta(S(\xi),S(y))]$.
The resulting ABC posterior, proportional to $\pi_0(x)L_{\ABC}(x)$, approximates the exact posterior, proportional to $\pi_0(x)p(y\mid x)$, and has two sources of error.
First, it conditions on the summaries $S(y)$ instead of the full data $y$, which loses information unless $S$ is sufficient.
Second, the kernel also gives weight to simulated summaries that are only close to $S(y)$; under mild conditions this error vanishes as $\delta\to0$.

A second example is a latent-variable model \cite{andrieu2009pseudomarginal}, in which unobserved random variables $\xi$ affect the data; for instance, repeated measurements on one individual may share an unobserved individual effect.
If $\xi$ were observed, the likelihood of $y$ given $(x,\xi)$ would be computable; we take this conditional likelihood as $\psi(x,\xi)$.
Averaging it over the unobserved $\xi\sim p(\cdot\mid x)$ gives the likelihood $L(x)$ in \eqref{eq:likelihood_expectation}, so here $L$ is the exact likelihood, with no ABC approximation.
More generally, $\xi$ can be the random inputs of any nonnegative unbiased estimator of $L(x)$.

By contrast, for a model $y=G(x)+\varepsilon$ with a deterministic forward map $G$ and additive noise with density $\rho$, independent of $x$,
the likelihood $L(x)=\rho(y-G(x))$ can be evaluated directly whenever $G$ can, and no auxiliary simulation is needed.

A direct Monte Carlo approximation of \eqref{eq:likelihood_expectation} is
\begin{equation}
 \widehat L_m(x)
 =
 \frac1m\sum_{i=1}^{m}\psi(x,\xi_i),
 \qquad
 \xi_i\stackrel{\mathrm{i.i.d.}}{\sim}p(\cdot\mid x).
\end{equation}
For fixed $x$, this estimator is unbiased, $\E[\widehat L_m(x)\mid x]=L(x)$.
Its drawback is cost.
Each likelihood value requires $m$ simulations, and a posterior sampler needs the likelihood at many parameter values, for example at every proposed state of a Markov chain.
The total number of simulations is therefore $m$ times the number of likelihood evaluations, and $m$ must be large enough to control the noise in each estimate.

Representation \eqref{eq:likelihood_expectation} also allows posterior expectations to be estimated without a sampler.
Given a proposal density $q$, draw $Z_i\sim q$ and $\xi_i\sim p(\cdot\mid Z_i)$, $i=1,\ldots,M$, and set $v_i=\pi_0(Z_i)\psi(Z_i,\xi_i)/q(Z_i)$.
Since $\E_q[v_ig(Z_i)]=\mathcal Z\,\E_\pi[g(X)]$, the self-normalized estimate $\sum_iv_ig(Z_i)/\sum_iv_i$ converges to $\E_\pi[g(X)]$ \cite[Chapter~9]{owen2013mc}.
This yields weighted estimates of posterior expectations, not posterior samples.
Resampling from the weighted empirical distribution $\sum_i(v_i/\sum_jv_j)\delta_{Z_i}$ gives equally weighted samples, but only at the points $Z_1,\ldots,Z_M$, and the number of distinct points is governed by the effective sample size $(\sum_iv_i)^2/\sum_iv_i^2$, which typically collapses as the dimension grows or as the posterior concentrates relative to $q$ \cite{agapiou2017importance}.

The sampler in Section~\ref{sec:methods} instead outputs unweighted samples and requires $L$ only through averages over parameter values.
By \eqref{eq:likelihood_expectation}, such averages are joint expectations over the parameter and the simulator output, and the simulations can be generated once by importance sampling and reused throughout the sampler.

\section{Likelihood-free SFS methods}
\label{sec:methods}

Section~\ref{sec:sfs} recalls SFS and its Euler--Maruyama discretization, in which the only quantity that cannot be computed directly is the drift.
Section~\ref{sec:joint_drift} writes the drift as a ratio of joint expectations over Gaussian variables and simulator outputs.
Sections~\ref{sec:numerical_method} and~\ref{sec:fixed_proposal_method} give two estimators of this ratio: \emph{likelihood-free SFS} draws new samples at each step, and \emph{likelihood-free SFS with sample reuse} uses importance sampling so that the samples are generated once and reused across all time steps and trajectories.

\subsection{Schr\"odinger--F\"ollmer sampling and its time discretization}
\label{sec:sfs}

Let
\begin{equation}
 \varphi_s(z):=(2\pi s)^{-d/2}\exp\!\left(-\frac{\|z\|^2}{2s}\right),
 \qquad s>0,\qquad \varphi:=\varphi_1,
\end{equation}
denote the densities of $\mathcal N(0,sI_d)$, where $I_d$ is the $d\times d$ identity matrix. For functions $g$ for which the integral below is finite, define
\begin{equation}
 f=\frac{\pi}{\varphi},\qquad
 Q_sg(x)=\int_{\R^d}g(z)\varphi_s(z-x)\,\mathrm dz.
 \label{eq:density_ratio}
\end{equation}
Thus $f$ is the target density relative to the Gaussian reference $\varphi$, and $Q_s$ is the heat semigroup.
With $B_t$ a standard $d$-dimensional Brownian motion, SFS \cite{huang2021sfs,jiao2021convergence} simulates
\begin{equation}
 \mathrm dX_t=b(X_t,t)\,\mathrm dt+\mathrm dB_t,
 \qquad X_0=0,\qquad b(x,t)=\nabla_x\log Q_{1-t}f(x),
 \label{eq:sfs_sde}
\end{equation}
for $0\le t<1$.

To see why $X_1\sim\pi$, let $\mathbb P_0$ be the law of a standard Brownian motion $(B_t)_{0\le t\le1}$ with filtration $(\mathcal F_t)$, and define $\mathbb P_*$ by
\begin{equation}
 \frac{\mathrm d\mathbb P_*}{\mathrm d\mathbb P_0}=f(B_1).
 \label{eq:sfs_change_of_measure}
\end{equation}
Since $\E_{\mathbb P_0}[f(B_1)]=\int f\varphi=\int\pi=1$, $\mathbb P_*$ is a probability measure, and under $\mathbb P_*$ the endpoint $B_1$ has density $f\varphi=\pi$.
By the Markov property, the density process is
\[
 M_t=\E_{\mathbb P_0}[f(B_1)\mid\mathcal F_t]=h(B_t,t),
 \qquad h(x,t)=Q_{1-t}f(x),\quad 0\le t<1.
\]
The function $h$ is positive and solves the backward heat equation $\partial_th+\frac12\Delta h=0$, so It\^o's formula gives
\[
 \mathrm dM_t=\nabla_xh(B_t,t)^\top\mathrm dB_t=M_t\,b(B_t,t)^\top\mathrm dB_t,
 \qquad b=\nabla_x\log h.
\]
By Girsanov's theorem \cite[Section~3.5]{karatzas}, $B_t-\int_0^tb(B_s,s)\,\mathrm ds$ is a Brownian motion under $\mathbb P_*$; that is, under $\mathbb P_*$ the process $B$ solves \eqref{eq:sfs_sde}.
Hence the solution of \eqref{eq:sfs_sde} has path law $\mathbb P_*$ and, in particular, $X_1\sim\pi$.
Making this rigorous up to $t=1$ requires regularity of $f$; Appendix~\ref{sec:proof_fresh} does so for the ABC targets of Section~\ref{sec:fresh_error}.

The drift has an explicit form.
If $f(z)\varphi_{1-t}(z-x)$ and its $x$-gradient are dominated by an integrable function of $z$, locally uniformly in $x$, then differentiating the Gaussian kernel in $Q_{1-t}f(x)$ with respect to $x$ gives
\begin{equation}
 b(x,t)=\frac{\E_Z[f(Z)(Z-x)]}{(1-t)\E_Z[f(Z)]},
 \qquad Z\sim\mathcal N(x,(1-t)I_d).
 \label{eq:intro_drift}
\end{equation}

To simulate \eqref{eq:sfs_sde}, choose an integer $K\ge1$ and a time grid
\begin{equation}
 0=t_0<\cdots<t_K=T\le1,\qquad
 \Delta t_k=t_{k+1}-t_k,\quad k=0,\ldots,K-1.
 \label{eq:time_grid}
\end{equation}
The uniform grid has $t_k=kT/K$ and $\Delta t_k=T/K$. Given an estimator $\widehat b$ of the drift, the Euler--Maruyama scheme has the iteration
\begin{equation}
 \begin{aligned}
 \widetilde X_{t_{k+1}}
 =\widetilde X_{t_k}+\Delta t_k\,\widehat b(\widetilde X_{t_k},t_k)
 +\sqrt{\Delta t_k}\,\eta_{k+1},\quad
 \widetilde X_0=0,\qquad
 \eta_{k+1}\overset{\mathrm{i.i.d.}}{\sim}\mathcal N(0,I_d)
 \end{aligned}
 \label{eq:euler_local}
\end{equation}
for $ k = 0,1,\dots,K-1$. The output is $\widetilde X_T$; taking $T=1$ targets the posterior $\pi$.

The scheme thus requires the drift at every visited state and time.
Formula \eqref{eq:intro_drift} involves no derivatives of the likelihood, but it requires values of $L$ through $f$, which are not available in our setting.
The rest of this section addresses the estimation of $b$.

\subsection{Likelihood-free representation of the drift}
\label{sec:joint_drift}

To remove $L$ from \eqref{eq:intro_drift}, we use the representation \eqref{eq:likelihood_expectation}. Draw
\begin{equation}
 Z\sim\mathcal N(x,(1-t)I_d),\qquad
 \xi\mid Z\sim p(\cdot\mid Z),
 \label{eq:joint_sampling}
\end{equation}
and define the computable weight
\begin{equation}
 W(z,\xi)=\frac{\pi_0(z)}{\varphi(z)}\psi(z,\xi).
 \label{eq:joint_weight}
\end{equation}
For fixed $z$, the likelihood representation gives
\begin{equation}
 \E_{\xi\mid z}[W(z,\xi)]
 =\frac{\pi_0(z)L(z)}{\varphi(z)}=\mathcal Z f(z).
 \label{eq:joint_weight_mean}
\end{equation}
By iterated expectation,
\begin{align*}
 \E_{Z,\xi}[W(Z,\xi)]
 &=\E_Z\!\left[\frac{\pi_0(Z)}{\varphi(Z)}
      \E_{\xi\mid Z}[\psi(Z,\xi)]\right]
 =\mathcal Z\E_Z[f(Z)],\\
 \E_{Z,\xi}[W(Z,\xi)(Z-x)]
 &=\E_Z\!\left[\frac{\pi_0(Z)}{\varphi(Z)}(Z-x)
      \E_{\xi\mid Z}[\psi(Z,\xi)]\right]
 =\mathcal Z\E_Z[f(Z)(Z-x)].
\end{align*}
Substituting these identities into \eqref{eq:intro_drift}, the constant $\mathcal Z$ cancels and
\begin{equation}
 b(x,t)=\frac{\E_{Z,\xi}[W(Z,\xi)(Z-x)]}
 {(1-t)\E_{Z,\xi}[W(Z,\xi)]}.
 \label{eq:problem_joint_drift}
\end{equation}
Both expectations are taken under the joint law \eqref{eq:joint_sampling}.
We require the denominator expectation $C(x,t)=\E_{Z,\xi}[W(Z,\xi)]$ to be finite and strictly positive, and
$\E[W(Z,\xi)\|Z-x\|]<\infty$ for the numerator.

\subsection{Likelihood-free SFS}
\label{sec:numerical_method}

The most direct estimator of \eqref{eq:problem_joint_drift} draws $m$ independent pairs $(Z_i,\xi_i)$ according to \eqref{eq:joint_sampling} and sets
\begin{equation}
 \widehat b_m(x,t)
 =\frac{m^{-1}\sum_{i=1}^m W(Z_i,\xi_i)(Z_i-x)}
 {(1-t)m^{-1}\sum_{i=1}^m W(Z_i,\xi_i)}.
 \label{eq:mc_local_problem}
\end{equation}
This is a self-normalized ratio estimator \cite[Section~9.2]{owen2013mc}: the numerator and denominator use the same pairs.
Each $Z_i$ is paired with a single draw $\xi_i\sim p(\cdot\mid Z_i)$.
Using $\widehat b=\widehat b_m$ in \eqref{eq:euler_local}, with new pairs generated at each step according to \eqref{eq:joint_sampling} at $(x,t)=(\widetilde X_{t_k},t_k)$, independently of future Brownian increments, gives \emph{likelihood-free SFS}.

Each drift evaluation uses $m$ evaluations of $\psi$, so $N$ trajectories with $K$ steps use $NKm$ evaluations in total; one evaluation of $\psi$ may itself require several simulator calls, as in Section~\ref{sec:exp3}.

The expectations in \eqref{eq:problem_joint_drift} can also be approximated by QMC or RQMC, replacing the independent uniform inputs that generate $(Z_i,\xi_i)$ with low-discrepancy points or their randomizations. At fixed $(x,t)$, the first $d$ coordinates of a QMC point $u_i$ in the open unit cube give $(Z_i)_j=x_j+\sqrt{1-t}\,\Phi_{\mathrm N}^{-1}(u_{i,j})$, $j=1,\ldots,d$, where $\Phi_{\mathrm N}$ is the standard normal cumulative distribution function; the remaining coordinates generate $\xi_i$ conditionally on $Z_i$. Since $\Phi_{\mathrm N}^{-1}$ diverges at $0$ and $1$, the transformed integrands may be unbounded, and the classical Koksma--Hlawka bound need not apply. For such integrands, under suitable growth conditions on the integrands and their mixed derivatives, Owen~\cite{owen2006halton} and Ouyang et~al.~\cite{ouyang2024unbounded} obtain RQMC errors of order $O(m^{-1+\varepsilon})$ in expectation for every $\varepsilon>0$, compared with the MC rate $O(m^{-1/2})$ for square-integrable integrands.

\subsection{Likelihood-free SFS with sample reuse}
\label{sec:fixed_proposal_method}

In likelihood-free SFS, the Gaussian samples at step $k$ are drawn from $\mathcal N(\widetilde X_{t_k},(1-t_k)I_d)$, which depends on both the current state and the time, so new $Z_i$ and new simulator outputs $\xi_i\sim p(\cdot\mid Z_i)$ are needed at every step and for every trajectory.

To keep the samples fixed, we use importance sampling. Choose a probability density $q$ on $\R^d$, independent of $(x,t)$, such that $q(z)>0$ for almost every $z$ with $\pi_0(z)L(z)>0$.
Draw $Z\sim q$ and then $\xi\sim p(\cdot\mid Z)$.  Write $\E_q$ for expectation under $q(z)\,\mathrm dz\,p(\mathrm d\xi\mid z)$.

Let $g$ be a scalar or vector function with $\E_{Z,\xi}[W(Z,\xi)\|g(Z)\|]<\infty$, where the expectation is under \eqref{eq:joint_sampling}. Then
\begin{align}
 \E_{Z,\xi}[W(Z,\xi)g(Z)]
 &=\int \E_{\xi\mid z}[W(z,\xi)]g(z)
       \varphi_{1-t}(z-x)\,\mathrm dz\notag\\
 &=\int q(z)\E_{\xi\mid z}[W(z,\xi)]
       \frac{\varphi_{1-t}(z-x)}{q(z)}g(z)\,\mathrm dz\notag\\
 &=\E_q\!\left[W(Z,\xi)
       \frac{\varphi_{1-t}(Z-x)}{q(Z)}g(Z)\right].
 \label{eq:qbank_change_measure}
\end{align}
The sampling law under $\E_q$ no longer depends on the state or the time. Taking $g=1$ and $g(z)=z-x$ in \eqref{eq:qbank_change_measure} gives
\begin{equation}
 b(x,t)=\frac{\E_q\!\left[\dfrac{W(Z,\xi)}{q(Z)}\varphi_{1-t}(Z-x)(Z-x)\right]}
 {(1-t)\E_q\!\left[\dfrac{W(Z,\xi)}{q(Z)}\varphi_{1-t}(Z-x)\right]}.
 \label{eq:qbank_exact_drift}
\end{equation}

Draw $M$ pairs once, with $Z_i\sim q$ and $\xi_i\sim p(\cdot\mid Z_i)$, and write
\begin{equation}
 \mathcal B_M=\{(Z_i,\xi_i)\}_{i=1}^M.
 \label{eq:qbank_bank}
\end{equation}
We store the values $W(Z_i,\xi_i)/q(Z_i)$. Replacing the expectations in \eqref{eq:qbank_exact_drift} by sample averages gives
\begin{equation}
 \widehat b_{M,q}(x,t)
 =\frac{\displaystyle\sum_{i=1}^M\frac{W(Z_i,\xi_i)}{q(Z_i)}\varphi_{1-t}(Z_i-x)(Z_i-x)}
 {(1-t)\displaystyle\sum_{i=1}^M\frac{W(Z_i,\xi_i)}{q(Z_i)}\varphi_{1-t}(Z_i-x)}.
 \label{eq:qbank_drift}
\end{equation}
At each $(x,t)$, only the Gaussian factors $\varphi_{1-t}(Z_i-x)$ must be recomputed. Using $\widehat b=\widehat b_{M,q}$ in \eqref{eq:euler_local}, with the same pairs at every step and for every trajectory, gives \emph{likelihood-free SFS with sample reuse}, which requires no further simulator evaluations.

For a fixed proposal $q$, likelihood-free SFS with sample reuse requires $M$ simulated pairs in total for $N$ trajectories of $K$ steps, compared with $NKm$ for likelihood-free SFS; this can substantially reduce computation time when simulator evaluations dominate the cost.

Section~\ref{sec:proposal_fitting} describes the proposals used in the experiments, and Sections~\ref{sec:fixed_sample_theory} and~\ref{sec:proposal_variance} analyze terminal convergence and pointwise drift estimation under sample reuse, respectively.

\subsection{Choosing the sampling distribution}
\label{sec:proposal_fitting}

Any proposal $q$ satisfying the support condition gives a consistent sampler, but the efficiency of sample reuse depends on $q$.
We compare three proposals of increasing sophistication, two of which are fitted to the posterior from a preliminary sample.
The experiments use coordinates in which the prior is standard Gaussian,
$\pi_0=\varphi$. Let $q_p$ be
a density for generating preliminary samples, positive wherever
$\varphi(z)L(z)>0$, and
draw $(Z_j^p,\xi_j^p)$, $j=1,\ldots,M_0$, with $Z_j^p\sim q_p$ and $\xi_j^p\mid Z_j^p\sim p(\cdot\mid Z_j^p)$.
Set
$\widetilde a(z,\xi)=\psi(z,\xi)\varphi(z)/q_p(z)$,
with sample values and normalized weights
\begin{equation}
 \widetilde a_j=\widetilde a(Z_j^p,\xi_j^p),
 \qquad
 a_j=\frac{\widetilde a_j}
 {\sum_{\ell=1}^{M_0}\widetilde a_\ell}.
 \label{eq:qbank_pilot_weights}
\end{equation}
These weights estimate posterior moments, since for integrable $g$
\begin{equation}
 \mathbb E_{q_p}[\widetilde a\,g(Z)]
 =
 \int\varphi(z)L(z)g(z)\,\mathrm dz.
 \label{eq:qbank_pilot_identity}
\end{equation}
The proposals use the fitted moments
\begin{equation}
 \widehat\mu=\sum_{j=1}^{M_0}a_jZ_j^p,
 \qquad
 \widehat\Sigma
 =
 \frac{\sum_{j=1}^{M_0}
 a_j(Z_j^p-\widehat\mu)(Z_j^p-\widehat\mu)^\top}
 {1-\sum_{j=1}^{M_0}a_j^2}.
 \label{eq:qbank_pilot_moments}
\end{equation}

Writing $\mathcal N(\mu,\Sigma)$ also for the corresponding density,
we use the proposals
\begin{equation}
 \begin{aligned}
 q_1&=\mathcal N(0,I_d),\\
 q_2&=\mathcal N\!\left(
       \widehat\mu,1.25\operatorname{diag}(\widehat\Sigma)
       \right),\\
 q_3&=0.95q_2+0.05q_1.
 \end{aligned}
 \label{eq:q123}
\end{equation}
Here $\operatorname{diag}(\widehat\Sigma)$ is the diagonal part of $\widehat\Sigma$.
The prior $q_1$ needs no fitting, but when the posterior is concentrated, few of its samples fall where the posterior has mass.
By \eqref{eq:joint_weight_mean}, the integrands in \eqref{eq:qbank_exact_drift} are, on average over $\xi$, proportional to $f(z)=\pi(z)/\varphi(z)$ and are therefore large where the posterior has mass; $q_2$ places its samples there by using the fitted posterior mean and variances.
The factor $1.25$ makes $q_2$ slightly wider than the posterior, because one fixed proposal must serve all states and times, and importance weights are more stable when the proposal is wider than the region it covers.
Finally, $q_3$ is a defensive mixture \cite{hesterberg1995,owen2000mixture}: since $q_3(z)\ge0.05q_1(z)$, its importance weights are at most $20$ times those under $q_1$, which protects against regions missed by the fitted Gaussian.
Section~\ref{sec:proposal_variance} analyzes how the proposal affects the drift-estimation variance.

In practice, a fitted variance that is nearly zero can be replaced by a small positive lower bound ($10^{-6}$ in Section~\ref{sec:exp3}).
Once fitted, the proposal is held fixed, and the samples used for drift estimation are generated independently of those used in fitting.

The local effective sample size (ESS) measures the concentration of the weights in \eqref{eq:qbank_drift} at a given state and time. When their sum is positive, it is defined by
\begin{equation}
 \operatorname{ESS}(x,t)
 =
 \frac{\left(\displaystyle\sum_{i=1}^M\frac{W(Z_i,\xi_i)}{q(Z_i)}\varphi_{1-t}(Z_i-x)\right)^2}
 {\displaystyle\sum_{i=1}^M\left(\frac{W(Z_i,\xi_i)}{q(Z_i)}\varphi_{1-t}(Z_i-x)\right)^2}.
 \label{eq:qbank_ess}
\end{equation}
It lies between $1$ and $M$ and does not count independent posterior outputs. Near the terminal time it can approach one even for the exact fixed-sample diffusion; Remark~\ref{rem:reuse_ess} relates it to the terminal importance weights.

\section{Error and convergence analysis}
\label{sec:theory}

We measure the error in the terminal distributions of the two methods of Section~\ref{sec:methods} in the quadratic Wasserstein distance $W_2$.
For likelihood-free SFS, Theorem~\ref{thm:path_error} separates time-discretization and drift-estimation errors.
For likelihood-free SFS with sample reuse, Theorem~\ref{thm:reuse_posterior_terminal} first identifies the fixed-sample terminal distribution and then proves convergence to the posterior.
The pointwise central limit theorem in Section~\ref{sec:proposal_variance} describes the effect of the proposal on drift estimation.

\subsection{Terminal error for likelihood-free SFS}
\label{sec:fresh_error}

We consider ABC in this subsection, i.e., $L=L_{\ABC}$ and $\psi(x,\xi)=K_\delta(S(\xi),S(y))$; write $\pi_{\ABC}$ for the resulting posterior.
Set $V=-\log f$, where $f=\pi_{\ABC}/\varphi$ as in \eqref{eq:density_ratio}.
\begin{assumption}
\label{ass:density_ratio_curvature}
The potential $V$ belongs to $C^2(\R^d)$, and there are constants $0\le\kappa<1$, $\lambda_+\ge0$, and $B_0\ge0$ such that, for all $x\in\R^d$,
\begin{equation}
 -\kappa I_d\preceq\nabla^2V(x)\preceq \lambda_+ I_d,\qquad
 \|\nabla V(0)\|\le B_0\sqrt d.
 \label{eq:Vcurvature}
\end{equation}
\end{assumption}

Since $-\log\pi_{\ABC}(x)=V(x)+\|x\|^2/2+\mathrm{constant}$ and $\kappa<1$, this assumption makes the target strongly log-concave.
It also yields the spatial Lipschitz and time-increment estimates for the drift used in the proof of the terminal error bound.
Appendix~\ref{sec:abc_verification} gives sufficient conditions in terms of the ABC likelihood and the prior.

Let $\widetilde X$ be likelihood-free SFS (Section~\ref{sec:numerical_method}) on the uniform grid \eqref{eq:time_grid} with $T=1$, $h=1/K$, and $t_k=kh$.
Let $\mathcal F_{t_k}$ be the $\sigma$-field generated by the Brownian motion up to time $t_k$ and all auxiliary samples generated before step $k$.
We control the drift-estimation error through a nonnegative function $\rho$ such that, almost surely for $k=0,\ldots,K-1$,
\begin{equation}
 \E\!\left[\|\widehat b_m(\widetilde X_{t_k},t_k)
 -b(\widetilde X_{t_k},t_k)\|^2\mid\mathcal F_{t_k}\right]
 \le\frac{\rho(t_k)d}{m},
 \label{eq:conditional_mse}
\end{equation}
and set
\begin{equation}
 \Lambda_h=h\sum_{k=0}^{K-1}\rho(t_k).
 \label{eq:integrated_local_error}
\end{equation}

\begin{theorem}
\label{thm:path_error}
Suppose that Assumption~\ref{ass:density_ratio_curvature} holds, that \eqref{eq:conditional_mse} holds, and that $\Lambda_h$ is finite for the chosen grid. Then
\begin{equation}
 W_2\!\left(\operatorname{Law}(\widetilde X_1),\pi_{\ABC}\right)
  \le C_{\mathrm{path}}\sqrt{dh}
  +C_{\mathrm{path}}\sqrt{\frac{d\Lambda_h}{m}},
 \label{eq:path_w2}
\end{equation}
where $C_{\mathrm{path}}$ depends only on $B_0$, $\lambda_+$, and $\kappa$.
\end{theorem}

The two terms in \eqref{eq:path_w2} are the time-discretization error and the accumulated drift-estimation error.
The proof in Appendix~\ref{sec:proof_fresh} couples the exact and numerical paths; it also shows that the fourth-moment conditions \eqref{eq:local_ratio_moments} are sufficient for \eqref{eq:conditional_mse}.
No bound on $\Lambda_h$ uniform in $h$ is required. For fixed $d$, the bound tends to zero whenever $h\to0$ and $\Lambda_h/m\to0$.
For example, when $f=1$ and $W=1$, the exact drift is zero, \eqref{eq:mc_local_problem} has mean-square error $d/[m(1-t)]$. Taking $\rho(t)=(1-t)^{-1}$ gives
\[
 \Lambda_h=h\sum_{k=0}^{K-1}\frac1{1-kh}
 =\sum_{j=1}^K\frac1j\le1+\log K.
\]
Thus \eqref{eq:path_w2} gives an error bound of order $O(\sqrt{dh}+\sqrt{d(1+\log(1/h))/m})$, which tends to zero for fixed $d$ if $h\to0$ and $\log(1/h)/m\to0$.

Theorem~\ref{thm:path_error} measures the error relative to the ABC posterior $\pi_{\ABC}$.
Relative to the posterior given the observed summary $S(y)$, the triangle inequality adds the ABC approximation error, which depends on the kernel, the summary, and the model, and vanishes as $\delta\to0$ under suitable conditions \cite{barber2015abc,frazier2018abc}; relative to the full-data posterior, a further error due to summarization appears.
Since $C_{\mathrm{path}}$ and $\Lambda_h$ may depend on $\delta$, a joint limit in $h$, $m$, and $\delta$ would require control of this dependence.

\subsection{Terminal convergence with sample reuse}
\label{sec:fixed_sample_theory}
\label{sec:reuse_terminal}

We now return to the general representation \eqref{eq:likelihood_expectation}; $f$, $W$, $b$, and $C$ refer to the corresponding target $\pi$.
For the samples $\mathcal B_M$ in \eqref{eq:qbank_bank}, define the importance weights and the weighted empirical measure
\begin{equation}
 \begin{aligned}
 c_i&=\frac{W(Z_i,\xi_i)}{q(Z_i)},\qquad
 v_i=c_i\varphi(Z_i)=\frac{\pi_0(Z_i)\psi(Z_i,\xi_i)}{q(Z_i)},\\
 a_{i,M}&=\frac{v_i}{\sum_{j=1}^M v_j},\qquad
 \nu_M=\sum_{i=1}^M a_{i,M}\delta_{Z_i},
 \end{aligned}
 \label{eq:reuse_terminal_measure}
\end{equation}
whenever $\sum_j v_j>0$; here $\delta_z$ denotes the Dirac measure at $z$.

Let $\widetilde X^{M,h}$ denote the Euler scheme \eqref{eq:euler_local} with drift $\widehat b_{M,q}$ on a deterministic grid \eqref{eq:time_grid} with $T=1$, driven by Brownian increments independent of the samples, and set $h=\max_k\Delta t_k$.
The exact diffusion with this drift, started at zero, has terminal distribution $\nu_M$ (Theorem~\ref{thm:reuse_posterior_terminal}(i) below), so the error splits as
\begin{equation}
 \begin{aligned}
 W_2\!\left(\mathcal L(\widetilde X_1^{M,h}\mid\mathcal B_M),\pi\right)
 \le
 W_2\!\left(\mathcal L(\widetilde X_1^{M,h}\mid\mathcal B_M),\nu_M\right)
 +W_2(\nu_M,\pi).
 \end{aligned}
 \label{eq:reuse_error_decomposition}
\end{equation}
The first term is the time-discretization error for fixed samples, and the second is the importance-sampling error.
For the large-sample limit, let $q$ be fixed and satisfy the support condition of Section~\ref{sec:fixed_proposal_method}, let $(Z_i,\xi_i)_{i\ge1}$ be an iid sequence with joint law $q(z)\,\mathrm dz\,p(\mathrm d\xi\mid z)$, and let $\mathcal B_M$ consist of its first $M$ pairs.

\begin{theorem}
\label{thm:reuse_posterior_terminal}
\emph{(i) Fixed samples.}
Suppose that $c_i\in[0,\infty)$ for all $i$ and $\sum_iv_i>0$.
Then the diffusion with drift $\widehat b_{M,q}$, started at zero, has a unique strong solution on $[0,1)$ that extends continuously to $t=1$, its terminal law given $\mathcal B_M$ is $\nu_M$, and, as $h\downarrow0$,
\begin{equation}
 W_2\!\left(\mathcal L(\widetilde X_1^{M,h}\mid\mathcal B_M),\nu_M\right)
 \longrightarrow0.
 \label{eq:reuse_euler_terminal}
\end{equation}

\emph{(ii) Increasing sample size.}
Suppose that $0<\mathcal Z<\infty$ and
\begin{equation}
 \int_{\R^d}\|z\|^2\pi(z)\,\mathrm dz<\infty.
 \label{eq:reuse_posterior_moment}
\end{equation}
Then, almost surely, $\sum_{i=1}^Mv_i>0$ for all sufficiently large $M$, and
\begin{equation}
 W_2(\nu_M,\pi)\longrightarrow0\quad\text{almost surely}.
 \label{eq:reuse_empirical_w2}
\end{equation}
\begin{equation}
 \lim_{M\to\infty}\lim_{h\downarrow0}
 W_2\!\left(\mathcal L(\widetilde X_1^{M,h}\mid\mathcal B_M),\pi\right)=0
 \quad\text{almost surely}.
 \label{eq:reuse_iterated_terminal}
\end{equation}
Moreover, there is a deterministic sequence of uniform grids with $h_M=1/K_M\downarrow0$ such that
\begin{equation}
 W_2\!\left(\mathcal L(\widetilde X_1^{M,h_M}\mid\mathcal B_M),\pi\right)
 \longrightarrow0\quad\text{almost surely}.
 \label{eq:reuse_diagonal_terminal}
\end{equation}
\end{theorem}

Thus both terms in \eqref{eq:reuse_error_decomposition} vanish, the first as $h\downarrow0$ for fixed samples and the second as $M\to\infty$.
The proof in Appendix~\ref{sec:proof_reuse} identifies the fixed-sample diffusion, controls its Euler approximation up to time one, and applies the strong law of large numbers to the weighted measure $\nu_M$.

\begin{corollary}[Unconditional terminal convergence]
\label{cor:reuse_unconditional_terminal}
Under the assumptions of part~(ii) of Theorem~\ref{thm:reuse_posterior_terminal}, suppose in addition that $\int\|z\|^2q(z)\,\mathrm dz<\infty$ and $v_1>0$ almost surely. Then
\[
 \lim_{M\to\infty}\lim_{h\downarrow0}
 W_2\!\left(\mathcal L(\widetilde X_1^{M,h}),\pi\right)=0.
\]
Moreover, the deterministic uniform grids in \eqref{eq:reuse_diagonal_terminal} can be chosen such that $W_2(\mathcal L(\widetilde X_1^{M,h_M}),\pi)\to0$.
\end{corollary}

The additional moment assumptions allow averaging over the sample set; the proof is in Appendix~\ref{sec:proof_reuse_unconditional}.
For a fitted proposal, the conditional results hold given the preliminary samples, provided the pairs used for drift estimation are conditionally iid and the assumptions hold for the fitted $q$; unconditional statements would require moment control over the fitting step.
RQMC samples and the input clipping used in Section~\ref{sec:exp3} are not covered; they would require separate analyses of the integration rule and of the resulting target approximation, respectively.

\subsection{Pointwise drift error and proposal choice}
\label{sec:proposal_variance}

Theorem~\ref{thm:reuse_posterior_terminal} holds for any proposal satisfying the support condition, but it does not show how the proposal affects accuracy.
This subsection addresses that question at a single state and time.
Theorem~\ref{thm:qbank_pointwise} shows that the reused drift estimator is consistent and asymptotically normal, and its asymptotic covariance then serves as a criterion for comparing proposals.
The main conclusions are that the variance-optimal proposal depends on the state and time, so that any fixed proposal is a compromise, and that the defensive mixture $q_3$ is never much worse than the prior $q_1$; this motivates the proposals of Section~\ref{sec:proposal_fitting}.
The analysis uses standard limit theory for self-normalized importance sampling \cite[Chapter~9]{owen2013mc}.
We abbreviate the weight by
\begin{equation}
 \Omega_q(z,\xi;x,t)=\frac{W(z,\xi)}{q(z)}\varphi_{1-t}(z-x).
 \label{eq:omega_q}
\end{equation}
At fixed $(x,t)$, write $\Omega_q=\Omega_q(Z,\xi;x,t)$. Identity \eqref{eq:qbank_change_measure} gives $\E_q\Omega_q=C(x,t)$, with $C$ defined in Section~\ref{sec:joint_drift}. Set
\begin{equation}
 \bar z(x,t)=x+(1-t)b(x,t).
 \label{eq:qbank_bridge_mean}
\end{equation}
By \eqref{eq:qbank_exact_drift}, $\bar z(x,t)=\E_q[\Omega_q Z]/C(x,t)$.
As in Section~\ref{sec:fixed_sample_theory}, the pairs in $\mathcal B_M$ are iid with law $q(z)\,\mathrm dz\,p(\mathrm d\xi\mid z)$, and $q$ satisfies the support condition of Section~\ref{sec:fixed_proposal_method}.

\begin{theorem}
\label{thm:qbank_pointwise}
Let $(x,t)$ with $t<1$ be fixed independently of $\mathcal B_M$. Suppose that $0<C(x,t)<\infty$ and
\begin{equation}
 \mathbb E_q\left[
 \Omega_q^2
 \left\|\frac{Z-x}{1-t}-b(x,t)\right\|^2
 \right]<\infty.
 \label{eq:qbank_centered_moment}
\end{equation}
Then, almost surely, $\widehat b_{M,q}(x,t)$ is well defined for all sufficiently large $M$, and
\begin{align}
 \widehat b_{M,q}(x,t)&\longrightarrow b(x,t)
 \quad\text{almost surely},
 \label{eq:qbank_slln}\\
 \sqrt M\{\widehat b_{M,q}(x,t)-b(x,t)\}
 &\Rightarrow\mathcal N(0,\Sigma_q(x,t)),
 \label{eq:qbank_clt}
\end{align}
where $\Rightarrow$ denotes convergence in distribution and
\begin{equation}
 \Sigma_q(x,t)
 =
 \frac{
 \mathbb E_q[
 \Omega_q^2(Z-\bar z)(Z-\bar z)^\top]}
 {(1-t)^2C(x,t)^2}.
 \label{eq:qbank_asymcov}
\end{equation}
\end{theorem}

The limiting covariance $\Sigma_q(x,t)$ may be singular.
The proof in Appendix~\ref{sec:proof_pointwise} applies the strong law of large numbers and the central limit theorem to a centered weighted sum.
These pointwise limits give no finite-sample mean-square bound and do not apply directly at numerical states, which depend on $\mathcal B_M$.

We now use $\operatorname{tr}\Sigma_q(x,t)$, the asymptotic mean-square error of the drift estimator at $(x,t)$, to compare proposals.

For fixed $(x,t)$, define
\begin{equation}
 \begin{aligned}
 \mathcal A_{x,t}(z)
 &=
 \varphi_{1-t}(z-x)^2
 \mathbb E_{\xi\mid z}\!\left[
 (W(z,\xi))^2\|z-\bar z(x,t)\|^2
 \right],\\
 \mathcal V_{x,t}(q)
 &=
 \int\frac{\mathcal A_{x,t}(z)}{q(z)}\,\mathrm dz.
 \end{aligned}
 \label{eq:qbank_variance_functional}
\end{equation}
Then
\begin{equation}
 \operatorname{tr}\Sigma_q(x,t)
 =
 \frac{\mathcal V_{x,t}(q)}
 {(1-t)^2C(x,t)^2}.
 \label{eq:qbank_trace}
\end{equation}
The denominator does not depend on $q$.
When $0<\int\sqrt{\mathcal A_{x,t}}<\infty$ and the support condition holds, the Cauchy--Schwarz inequality shows that the trace is minimized by
\begin{equation}
 q^*_{x,t}(z)
 =
 \frac{\sqrt{\mathcal A_{x,t}(z)}}
 {\int\sqrt{\mathcal A_{x,t}(u)}\,\mathrm du}.
 \label{eq:qbank_optimal_proposal}
\end{equation}
Since $q^*_{x,t}$ depends on $(x,t)$, on the unknown $\bar z(x,t)$, and on a conditional second moment of the simulator weight, a single fitted proposal cannot in general minimize this variance at all states and times.
A proposal fitted to the posterior, such as $q_2$, is a practical compromise, since the numerical states concentrate near the posterior mass as $t\to1$.

For the mixture $q_3$ in \eqref{eq:q123}, the bound $q_3(z)\ge0.05q_1(z)$ gives
\begin{equation}
 \mathcal V_{x,t}(q_3)
 \le20\,\mathcal V_{x,t}(q_1)
 \label{eq:qbank_defensive_bound}
\end{equation}
provided $\mathcal V_{x,t}(q_1)<\infty$. Thus, at every state and time, the asymptotic mean-square drift error under $q_3$ is at most $20$ times that under the prior $q_1$, whatever the quality of the fitted component.

These calculations are detailed in Appendix~\ref{sec:proposal_variance_details}.

\section{Numerical experiments}

\label{sec:experiments}

\rev{With sample reuse, trajectories driven by independent Brownian paths are conditionally independent given $\mathcal B_M$ but generally dependent unconditionally. Increasing $N$ therefore does not reduce the error due to $\mathcal B_M$; assessing its variability requires independent repetitions of the sample generation.}
In the tables and figures, MC--SFS and RQMC--SFS denote likelihood-free SFS whose drift samples are generated by MC and RQMC, respectively, and $q_j$--MC and $q_j$--RQMC denote likelihood-free SFS with sample reuse, proposal $q_j$ in \eqref{eq:q123}, and MC or RQMC samples.

\subsection{\texorpdfstring{\rev{Analytic Gaussian ABC benchmark}}{Analytic Gaussian ABC benchmark}}

\label{sec:exp1}

\rev{This model has a closed-form ABC posterior and SFS drift, which allows us to assess drift estimation, sample reuse, and Euler discretization, including for a concentrated posterior. Let $x\sim\mathcal N(0,I_2)$ and}
\begin{equation}
 \xi\mid x\sim\mathcal N(x,\sigma^2I_2),
 \qquad
 y=(2,-2)^\top,
 \qquad
 \sigma=0.5.
\end{equation}

\rev{We use the identity summary and the Gaussian ABC kernel}
\begin{equation}
 \rev{\psi}(x,\xi)
 =
 \exp\left[-\frac{\|\xi-y\|^2}{2\delta}\right],
 \qquad \delta=0.25.
 \label{eq:exp1_kernel}
\end{equation}

\rev{Here $\delta$ is a bandwidth on the variance scale. Since $\rev{\psi}(x,\xi)=(2\pi\delta)^{d/2}\varphi_\delta(\xi-y)$, Gaussian convolution gives}
\begin{align}
 \rev{L}_{\ABC}(x)
 &=
 (2\pi\delta)^{d/2}
 \int_{\mathbb R^d}
 \varphi_\delta(\xi-y)
 \varphi_{\sigma^2}(\xi-x)\,\mathrm d\xi
 \notag\\
 &=
 \left(\frac{\delta}{\delta+\sigma^2}\right)^{d/2}
 \exp\left[
 -\frac{\|x-y\|^2}{2(\delta+\sigma^2)}
 \right],
 \qquad d=2.
\end{align}

\rev{Set $v=\sigma^2+\delta=0.5$. Multiplying by the prior and completing the square gives}
\begin{equation}
 \rev{\pi}_{\ABC}=\mathcal N(\mu_\star,\Sigma_\star),
 \qquad
 \mu_\star=\frac{y}{1+v}=(4/3,-4/3)^\top,
 \qquad
 \Sigma_\star=\frac{v}{1+v}I_2=\tfrac13I_2.
\end{equation}

\rev{Since $\rev{f}=\rev{L}_{\ABC}/\rev{\mathcal Z}$, a second Gaussian convolution yields}
\begin{equation}
 Q_{1-t}\rev{f}(x)
 =
 c_{\mathrm G}(t)\exp\left[
 -\frac{\|x-y\|^2}{2(v+1-t)}
 \right],
 \label{eq:exp1_heat}
\end{equation}
\rev{where $c_{\mathrm G}(t)>0$ is independent of $x$. Hence}
\begin{equation}
 b(x,t)
 =
 \nabla_x\log Q_{1-t}\rev{f}(x)
 =
 \frac{y-x}{v+1-t}.
 \label{eq:exp1_true_drift}
\end{equation}

\rev{The density-ratio potential satisfies}
\begin{equation}
 V(x)=-\log \rev{f}(x)
 =\frac{\|x-y\|^2}{2v}+\mathrm{constant},
 \qquad
 \nabla^2V(x)=2I_2,
 \qquad
 \nabla V(0)=-2y.
 \label{eq:exp1_potential}
\end{equation}

\rev{Thus Assumption~\ref{ass:density_ratio_curvature} holds with $\kappa=0$, $\lambda_+=2$, and $B_0=4$, since $\|\nabla V(0)\|=4\sqrt2=B_0\sqrt d$. However, for every finite $m$, no deterministic $\rho$ satisfies \eqref{eq:conditional_mse} uniformly over the states $x$. Indeed, with $\tau=1-t$ and $G_i=(Z_i-x)/\sqrt\tau$, the estimator $\widehat b_m$ is a convex combination of $G_i/\sqrt\tau$, so}
\begin{revision}
\[
 \E\|\rev{\widehat b_m}(x,t)\|^2\le\frac{md}{\tau},
 \qquad
 \bigl(\E\|\rev{\widehat b_m}(x,t)-b(x,t)\|^2\bigr)^{1/2}
 \ge\left(\frac{\|y-x\|}{v+\tau}-\sqrt{\frac{md}{\tau}}\right)_+.
\]
\end{revision}

\noindent\rev{where $(u)_+=\max\{u,0\}$. The lower bound diverges as $\|x\|\to\infty$. Since every Euler state after the first step has full support, owing to its independent Gaussian increment, no finite deterministic $\rho(t_k)$ satisfies \eqref{eq:conditional_mse} at those steps. Theorem~\ref{thm:path_error} therefore does not directly cover this experiment, although this does not indicate that the algorithm fails. All comparisons target the ABC posterior with $\delta=0.25$ and do not test the limit $\delta\to0$.}

\rev{Let $(\widehat\mu_{\mathrm{out}},\widehat\Sigma_{\mathrm{out}})$ be the sample mean and covariance of the output, and $(\mu_{\mathrm{ref}},\Sigma_{\mathrm{ref}})$ the analytic reference moments, namely $(\mu_\star,\Sigma_\star)$ here and $(\mu_{100},\Sigma_{100})$ in the concentrated case below. Define}
\begin{equation}
 E_\mu
 =
 \frac{\|\widehat\mu_{\mathrm{out}}-\mu_{\mathrm{ref}}\|}
 {\sqrt{\operatorname{tr}\Sigma_{\mathrm{ref}}}},
 \qquad
 E_\Sigma
 =
 \frac{\|\widehat\Sigma_{\mathrm{out}}-\Sigma_{\mathrm{ref}}\|_F}
 {\|\Sigma_{\mathrm{ref}}\|_F}.
 \label{eq:exp1_moment_errors}
\end{equation}

\rev{When the full output covariance was recorded, we also report}
\begin{equation}
 \begin{split}
 W_{2,G}^2
 ={}&\|\widehat\mu_{\mathrm{out}}-\mu_{\mathrm{ref}}\|^2\\
 &+\operatorname{tr}\left[
 \widehat\Sigma_{\mathrm{out}}+\Sigma_{\mathrm{ref}}
 -2(\Sigma_{\mathrm{ref}}^{1/2}\widehat\Sigma_{\mathrm{out}}
 \Sigma_{\mathrm{ref}}^{1/2})^{1/2}
 \right],
 \end{split}
 \label{eq:gaussian_w2}
\end{equation}
\rev{where the matrix square roots are positive semidefinite. This is the squared $W_2$ distance between the Gaussian distributions with the estimated and the reference moments \cite{gelbrich1990}; for non-Gaussian output it does not measure the full distributional error.}

\rev{As a baseline, we ran the preconditioned Crank--Nicolson (pCN) sampler \cite{cotter2013pcn} with the closed-form ABC likelihood for $100000$ steps, discarding the first $20000$ and retaining every twentieth subsequent state. Because this baseline uses the closed-form likelihood, it is not an equal-cost, simulator-only comparison; the reference moments are taken from the closed-form posterior, not from the pCN output.}

\rev{For likelihood-free SFS, we used $N=1000$ outputs, $K=80$ steps, and $m\in\{64,128,256\}$ pairs per drift evaluation. The grid \eqref{eq:time_grid} had $T=1$ and}
\begin{revision}
\begin{equation}
 t_k=1-\left(1-\frac{k}{K}\right)^3,
 \qquad k=0,\ldots,K.
 \label{eq:exp1_local_time_grid}
\end{equation}
\end{revision}

\noindent\rev{so the steps shorten near time one, unlike the uniform grid of Theorem~\ref{thm:path_error}. Table~\ref{tab:exp1_local} reports the resulting posterior moment errors, averaged over five independent repetitions.}

\begin{table}[!tbp]
\centering
\caption{\rev{Posterior moment errors for likelihood-free SFS in}
the Gaussian ABC model}
\label{tab:exp1_local}
\begin{tabular}{lccc}
\toprule
Method & $m$ & $E_\mu$ & $E_\Sigma$\\
\midrule
MC--SFS &64&0.0949&0.1571\\
MC--SFS &128&0.0442&0.1085\\
MC--SFS &256&0.0284&0.0920\\
RQMC--SFS &64&0.0521&0.1147\\
RQMC--SFS &128&0.0312&0.0690\\
RQMC--SFS &256&0.0247&0.0678\\
pCN (closed-form $L_{\ABC}$) &---&0.0126&0.0284\\
\bottomrule
\end{tabular}
\end{table}

\begin{figure}[!tbp]
\centering
\includegraphics[width=.94\textwidth]{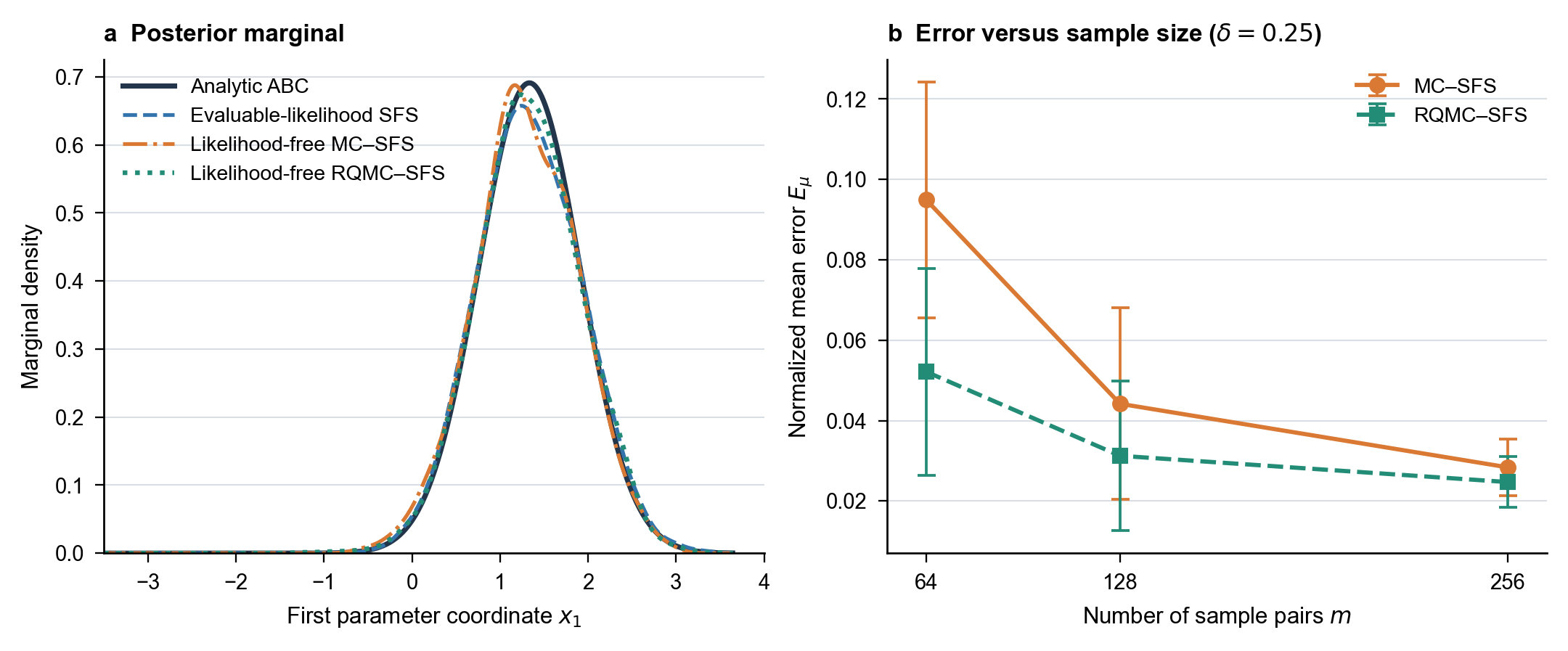}
\caption{Gaussian ABC validation with variance bandwidth $\delta=0.25$. (a) Analytic first-coordinate posterior density and kernel density estimates from one run per SFS method ($N=3000$, $K=80$, $m=256$). SFS with the closed-form likelihood uses the weights $L_{\ABC}(Z_i)$ at Gaussian samples $Z_i$, rather than the exact drift. (b) Mean normalized posterior-location error $E_\mu$ versus the sample size $m$ for likelihood-free MC--SFS and RQMC--SFS ($N=1000$, $K=80$). Points are means and error bars are sample standard deviations across five independent repetitions. Both panels use \eqref{eq:exp1_local_time_grid}.}
\label{fig:exp1_local}
\end{figure}

\rev{As Table~\ref{tab:exp1_local} and Fig.~\ref{fig:exp1_local}(b) show, both moment errors decreased with $m$, and RQMC gave lower errors than MC for every $m$; neither reached the accuracy of the pCN baseline, which uses the closed-form likelihood. With $K$ and $N$ fixed, these errors combine the drift-estimation error, the time-discretization error, and the Monte Carlo error due to the finite number $N$ of outputs. Increasing $m$ reduces only the drift-estimation error, so the comparison does not isolate a discretization or terminal convergence rate.}

\rev{Fig.~\ref{fig:exp1_local}(a) compares the analytic first-coordinate marginal $\mathcal N(4/3,1/3)$ with three separate SFS runs, each using $N=3000$, $K=80$, $m=256$, and \eqref{eq:exp1_local_time_grid}. In each run, the drift is a self-normalized weighted mean of $(Z_i-x)/(1-t)$ with $Z_i=x+\sqrt{1-t}\,\eta_i$. For comparison, SFS with the closed-form likelihood uses independent standard Gaussian $\eta_i$ and weights $\rev{L}_{\ABC}(Z_i)$; it still estimates the drift by Monte Carlo rather than using \eqref{eq:exp1_true_drift}. Likelihood-free MC--SFS uses independent standard Gaussian $\eta_i,\zeta_i\in\R^2$, one output $\xi_i=Z_i+\sigma\zeta_i$ per $Z_i$, and weights $\rev{\psi}(Z_i,\xi_i)$. RQMC--SFS uses the same weights with a four-dimensional Sobol rule, scrambled independently at each drift evaluation and mapped through the inverse normal distribution function. All Brownian increments are pseudorandom. The kernel density estimates use a Gaussian kernel with bandwidth $1.06\,sN^{-1/5}$, where $s$ is the sample standard deviation of the run. These runs are separate from the five-repetition comparison in panel~(b).}

\rev{For likelihood-free SFS with sample reuse, we generated $M\in\{1024,4096,16384\}$ pairs using scrambled Sobol points. The proposals follow \eqref{eq:q123}, with $M_0=4096$ independent preliminary samples for fitting $q_2$ and $q_3$. Each configuration used $N=1000$ trajectories on the uniform grid \eqref{eq:time_grid} with $T=1$ and $K=80$, and was repeated five times.}

\rev{In each repetition, the drift RMSE is the square root of the squared Euclidean error relative to the analytic drift \eqref{eq:exp1_true_drift}, averaged over all $N$ trajectories and all update times $t_0,\ldots,t_{K-1}$. For each $M$ and proposal, Table~\ref{tab:exp1_bank} reports the moment errors, $W_{2,G}$, this drift RMSE averaged over the five repetitions, and the number of simulator calls.}

\begin{table}[!tbp]
\centering
\small
\caption{Gaussian ABC: \rev{errors and simulator calls for likelihood-free SFS with sample reuse};
\rev{calls include proposal fitting where applicable}}
\label{tab:exp1_bank}
\begin{tabular}{llrrrrr}
\toprule
$M$&Proposal&$E_\mu$&$E_\Sigma$&$W_{2,G}$&Drift RMSE&Calls\\
\midrule
1024&$q_1$&0.0813&0.1119&0.0848&1.1696&1024\\
1024&$q_2$&0.0637&0.1145&0.0711&0.5813&5120\\
1024&$q_3$&0.0506&0.0651&0.0518&0.5446&5120\\
\midrule
4096&$q_1$&0.0496&0.0573&0.0478&0.6492&4096\\
4096&$q_2$&0.0265&0.0446&0.0302&0.2966&8192\\
4096&$q_3$&0.0211&0.0572&0.0299&0.3584&8192\\
\midrule
16384&$q_1$&0.0371&0.0430&0.0370&0.2882&16384\\
16384&$q_2$&0.0268&0.0435&0.0302&0.1708&20480\\
16384&$q_3$&0.0193&0.0669&0.0324&0.2250&20480\\
\bottomrule
\end{tabular}
\end{table}

\begin{figure}[!tbp]
\centering
\includegraphics[width=.94\textwidth]{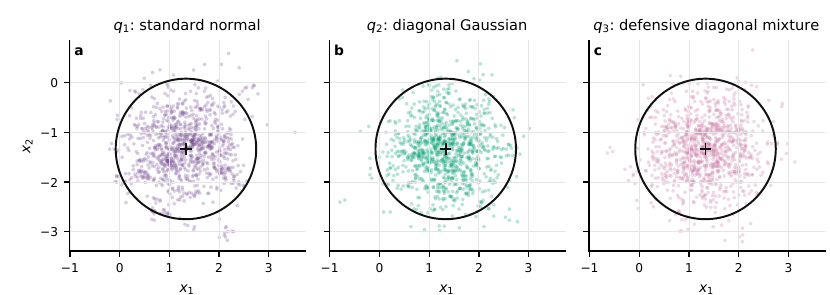}
\caption{Terminal samples for the three proposals at
$M=4096$. Black ellipses show the analytic posterior's
$95\%$ probability contour; crosses mark its mean.}
\label{fig:exp1_bank}
\end{figure}

\rev{At $M=4096$, $q_2$ and $q_3$ reduced $W_{2,G}$ from $0.0478$ to $0.0302$ and $0.0299$, respectively (about $37\%$), and the drift RMSE from $0.6492$ to $0.2966$ and $0.3584$. Increasing $M$ reduced the drift RMSE for every proposal, but the terminal errors did not decrease monotonically: for the fitted proposals, $W_{2,G}$ did not decrease from $M=4096$ to $M=16384$. Since the drift errors are measured along each method's own trajectories, they do not compare estimates at common states.}
Fig.~\ref{fig:exp1_bank} shows the corresponding terminal samples at $M=4096$ for the three proposals, together with the $95\%$ probability contour of the analytic posterior.

\rev{Likelihood-free SFS with $m=128$ used $NKm=10\,240\,000$ simulator calls, compared with $M_0+M=8192$ for sample reuse with $M=4096$ and either fitted proposal; $q_1$ requires no preliminary samples. These counts measure simulator savings, not reductions of the same size in arithmetic or runtime, since $\Omega_q(Z_i,\xi_i;x,t)$ must be recomputed at each state and time. Because the samples are scrambled Sobol points rather than iid draws, Theorems~\ref{thm:qbank_pointwise} and~\ref{thm:reuse_posterior_terminal} do not directly cover this experiment.}

\rev{To examine a concentrated posterior, we next let each simulator call generate $n=50$ independent bivariate observations $\xi_1,\ldots,\xi_n$, so that each simulated data set has $2n=100$ coordinates, while keeping two parameters and the two-dimensional summary $\bar\xi=n^{-1}\sum_{j=1}^n\xi_j$. With observed mean $y_0=(2,-2)^\top$,}
\begin{equation}
 \bar\xi\mid x\sim\mathcal N(x,\sigma^2I_2/n).
 \label{eq:exp1_summary_law}
\end{equation}

\rev{With the bandwidth $\delta_S=\delta/n=0.005$ for the summary, the variance $v=\sigma^2+\delta$ in the posterior formula is replaced by $(\sigma^2+\delta)/n=v/n$, which gives}
\begin{equation}
 \begin{aligned}
 \mu_{100}
 &=\frac{n}{n+v}y_0
   =(1.980198,-1.980198)^\top,\\
 \Sigma_{100}
 &=\frac{v}{n+v}I_2
   =0.00990099I_2.
 \end{aligned}
 \label{eq:exp1_concentrated_target}
\end{equation}

\rev{The subscript $100$ refers to the $2n=100$ data coordinates, not to the number of parameters. We used $N=1000$, $m=128$, $M=M_0=4096$, and five repetitions on the uniform grid \eqref{eq:time_grid} with $T=1$ and $K=80$.}

\begin{figure}[!tbp]
\centering
\includegraphics[width=.92\textwidth]{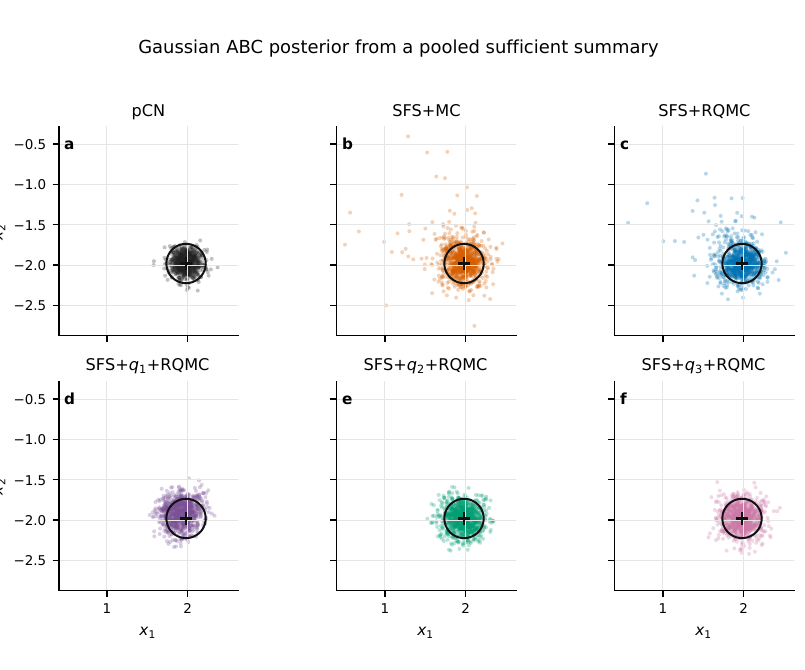}
\caption{Posterior projections for $n=50$ bivariate observations ($100$ data coordinates). Reusing samples drawn from fitted proposals improves centering, while the
uniform grid leaves visible variance inflation relative
to the closed-form $95\%$ ellipses.}
\label{fig:exp1_100d}
\end{figure}

Table~\ref{tab:exp1_100d} compares the posterior errors and simulator calls of all methods, and Fig.~\ref{fig:exp1_100d} shows the corresponding posterior projections against the closed-form $95\%$ ellipses.

\begin{table}[!tbp]
\centering
\small
\caption{Posterior errors for $n=50$ bivariate observations ($100$ data coordinates)
and the summary $\bar\xi\in\R^2$}
\label{tab:exp1_100d}
\begin{tabular}{lrrrr}
\toprule
Method&$E_\mu$&$E_\Sigma$&$W_{2,G}$&Calls\\
\midrule
pCN (closed-form $L_{\ABC}$)&0.0184&0.0333&0.0036&---\\
MC--SFS&0.3320&2.8019&0.1385&10\,240\,000\\
RQMC--SFS&0.2109&2.0244&0.1058&10\,240\,000\\
$q_1$--RQMC&0.6412&0.9805&0.1066&4096\\
$q_2$--RQMC&0.0741&0.8402&0.0512&8192\\
$q_3$--RQMC&0.0752&0.8352&0.0511&8192\\
\bottomrule
\end{tabular}
\end{table}

\rev{The fitted proposals improved the centering, but the relative covariance error remained about $0.84$. Part of this error is due to the time discretization. With $h=\Delta t_k=1/80$, the final Euler update is}
\begin{equation}
 \widetilde X_1=\widetilde X_1^{\mathrm{pre}}+\sqrt h\,\eta_K,
 \qquad
 \widetilde X_1^{\mathrm{pre}}=\widetilde X_{1-h}
   +h\widehat b(\widetilde X_{1-h},1-h),
 \label{eq:exp1_final_update}
\end{equation}
\rev{where $\widehat b$ is the drift estimator used and $\eta_K\sim\mathcal N(0,I_2)$ is independent of the preceding state, the samples, and the drift estimate. If second moments exist, then}
\begin{equation}
 \operatorname{Cov}(\widetilde X_1)
 =
 \operatorname{Cov}(\widetilde X_1^{\mathrm{pre}})+hI_2
 \succeq hI_2
 =
 0.0125I_2.
 \label{eq:exp1_terminal_covariance_floor}
\end{equation}

\rev{This lower bound exceeds $\Sigma_{100}=0.00990099I_2$, so on this grid the covariance of the Euler output cannot equal the target covariance, even with the exact drift. The bound identifies a discretization contribution that more reused samples cannot remove; it does not account for the entire observed covariance error. This experiment varies the posterior concentration, not the parameter dimension, and therefore does not test the dependence on $d$ in the theory.}

\subsection{\texorpdfstring{\rev{M/G/1 queueing model}}{M/G/1 queueing model}}

\label{sec:exp2}

\rev{The M/G/1 benchmark \cite{blum2010nonlinear} tests likelihood-free integration with a piecewise-smooth simulator and high-dimensional random input. For an initially empty, single-server, first-come-first-served queue, set $a_0=d_0=0$ and}
\begin{equation}
\begin{aligned}
 T_i^{\mathrm{svc}}&=\theta_1+\theta_2U_i,
 &U_i&\sim\mathrm{Unif}(0,1),\\
 T_i^{\mathrm{ia}}&=-\log(1-U_i')/\theta_3,
 &U_i'&\sim\mathrm{Unif}(0,1),\\
 a_i&=a_{i-1}+T_i^{\mathrm{ia}},
 &d_i&=\max(a_i,d_{i-1})+T_i^{\mathrm{svc}},\\
 Y_i&=d_i-d_{i-1}.
\end{aligned}
\label{eq:mg1_recursion}
\end{equation}

\rev{All uniforms are mutually independent, and $T_i^{\mathrm{svc}}$, $T_i^{\mathrm{ia}}$, $a_i$, $d_i$, and $Y_i$ denote the service, interarrival, arrival, departure, and interdeparture times, respectively. Service times are uniform on $(\theta_1,\theta_1+\theta_2)$, so $\theta_2$ is the interval width, and $\theta_3$ is the arrival rate. The data are 50 interdeparture times simulated at $\theta=(1,4,0.2)$, and the priors are independent, with $\theta_1,\theta_2\sim\mathrm{Unif}(0,10)$ and $\theta_3\sim\mathrm{Unif}(0,1/3)$. The likelihood is that of this finite sequence from an initially empty queue, not of a stationary queue. SFS runs in standard Gaussian coordinates $x\in\R^3$, mapped to the original parameters $\theta$ by}
\begin{equation}
 \theta=F(x)
 =
 \bigl(10\Phi_{\mathrm N}(x_1),10\Phi_{\mathrm N}(x_2),
 \Phi_{\mathrm N}(x_3)/3\bigr)^\top,
 \label{eq:mg1_transform}
\end{equation}

\rev{By the probability integral transform, $x\sim\mathcal N(0,I_3)$ yields the stated priors.}

\rev{We used the five-dimensional summary}
\begin{equation}
 S(Y)
 =
 \bigl(\log Q_0,\log Q_{0.25},\log Q_{0.5},
       \log Q_{0.75},\log Q_1\bigr)^\top,
 \label{eq:mg1_summary}
\end{equation}
\rev{where $Q_p$ is the empirical $p$-quantile of the 50 interdeparture times. The observed summary and componentwise scales were}
\begin{equation}
\begin{aligned}
 \rev{s_{\mathrm{obs}}}&=(0.0929,0.8333,1.4484,1.9773,3.1510)^\top,\\
 s_{\mathrm{sc}}&=(0.1049,0.1336,0.1006,0.1893,0.2918)^\top,
 \qquad D_{\mathrm{sc}}=\operatorname{diag}(s_{\mathrm{sc}}).
\end{aligned}
\label{eq:mg1_summary_scale}
\end{equation}

\rev{Let $\omega=(U_1,U'_1,\ldots,U_{50},U'_{50})$ be the 100-dimensional simulator input, playing the role of $\xi$ in Section~\ref{sec:problem}. The ABC weight is}
\begin{equation}
 \rev{\psi}(x,\omega)
 =
 \exp\left[
 -\frac{
 \|D_{\mathrm{sc}}^{-1}\{S(Y(F(x),\omega))-\rev{s_{\mathrm{obs}}}\}\|^2}
 {2\delta}
 \right].
 \label{eq:mg1_abc_weight}
\end{equation}

\rev{Thus $\rev{L}_{\ABC}(x)=\mathbb E_\omega[\rev{\psi}(x,\omega)]$, with parameter dimension $d=3$ and summary dimension $r=5$. The bandwidth was calibrated on $65536$ RQMC points from the prior so that the ESS of the ABC weights equals $5\%$ of the number of points, which gives $\delta=7.211990$; the fractions $0.02$ and $0.10$ would give $\delta=3.307955$ and $15.028490$. All reported comparisons use $\delta=7.211990$. The ESS fraction measures weight concentration rather than an ABC acceptance rate; once chosen, the bandwidth and the scaling fix the ABC target.}

\rev{The maximum and quantile operations make the simulator output only piecewise smooth in $\theta$, although the integrated likelihood may still be smooth. Assumption~\ref{ass:density_ratio_curvature} and the moment conditions \eqref{eq:local_ratio_moments} have not been verified, so Theorem~\ref{thm:path_error} is not known to apply; the comparisons also do not test the ABC bandwidth limit or the full-data posterior.}

\rev{The numerical reference was computed from 524288 RQMC points drawn from the prior, retaining 32768 weighted samples. Two independent reference sets had ESS 26082 and 25944, a sliced Wasserstein discrepancy of $0.0734$ between them, and a classifier two-sample test accuracy near $0.5$. These checks indicate reproducibility but do not make the reference exact. Proposal fitting used $M_0=16384$ preliminary points with ESS $821.46$, giving}
\begin{equation}
 \begin{aligned}
 \widehat\mu
 &=(-1.28810,-0.06673,0.41174)^\top,\\
 \bigl((\widehat\Sigma_{q_2})_{11},
       (\widehat\Sigma_{q_2})_{22},
       (\widehat\Sigma_{q_2})_{33}\bigr)
 &=(0.06398,0.88107,0.46020).
 \end{aligned}
 \label{eq:mg1_fitted_proposal}
\end{equation}

\rev{Here $\widehat\mu$ is fitted as in Section~\ref{sec:proposal_fitting}, and $\widehat\Sigma_{q_2}$ is the proposal covariance, not a posterior covariance estimate; the proposals are defined in \eqref{eq:q123}. Each combination of proposal and $M$ used $N=384$ outputs, the uniform grid \eqref{eq:time_grid} with $T=1$ and $K=64$, and twelve repetitions. Error bars in the figures are $1.96$ standard errors of the mean; errors of the terminal distribution are reported as mean $\pm$ standard deviation. Each RQMC point has $3+100=103$ coordinates, or $104$ for $q_3$, which uses one more coordinate to select the mixture component; this is much larger than the parameter dimension $d=3$.}

\rev{Let $(\mu_{\mathrm{ref}}^\theta,\Sigma_{\mathrm{ref}}^\theta)$ be the numerical reference moments in the original coordinates $\theta=F(x)$, and set $D_{\mathrm{ref}}=\operatorname{diag}(\sqrt{(\Sigma_{\mathrm{ref}}^\theta)_{11}},\ldots,\sqrt{(\Sigma_{\mathrm{ref}}^\theta)_{dd}})$. Both sliced $W_2$ \cite{bonneel2015sliced} and $W_{2,G}$ in \eqref{eq:gaussian_w2} are computed in the standardized coordinates $D_{\mathrm{ref}}^{-1}(\theta-\mu_{\mathrm{ref}}^\theta)$ for both output and reference samples. Sliced $W_2$ uses 384 random projection directions with 384 samples per direction. For the output moments $(\widehat\mu_{\mathrm{out}}^\theta,\widehat\Sigma_{\mathrm{out}}^\theta)$, define}
\begin{revision}
\begin{equation}
 \begin{split}
 E_{\mu,\theta}
 &={}
 \frac{\|D_{\mathrm{ref}}^{-1}
 (\widehat\mu_{\mathrm{out}}^\theta
  -\mu_{\mathrm{ref}}^\theta)\|}{\sqrt d},\\
 E_{\Sigma,\theta}
 &={}
 \frac{\|D_{\mathrm{ref}}^{-1}
 (\widehat\Sigma_{\mathrm{out}}^\theta-\Sigma_{\mathrm{ref}}^\theta)
 D_{\mathrm{ref}}^{-1}\|_F}
 {\|D_{\mathrm{ref}}^{-1}\Sigma_{\mathrm{ref}}^\theta
 D_{\mathrm{ref}}^{-1}\|_F}.
 \end{split}
 \label{eq:mg1_mean_error}
\end{equation}
\end{revision}

\noindent\rev{Here $d=3$; this coordinatewise standardization differs from the normalization in Section~\ref{sec:exp1}. These projection and moment diagnostics differ from the full $W_2$ distance in Theorem~\ref{thm:path_error}; moreover, they concern the distribution of $\theta=F(x)$, whereas the theorem uses Gaussian coordinates.}

\rev{For the drift comparisons, we fixed six points $x_j$ in Gaussian coordinates and evaluated the drift at the states $\sqrt t\,x_j$ for}
\begin{equation}
 t\in\{0,0.25,0.5,0.75,0.9,0.975\}.
 \label{eq:mg1_query_times}
\end{equation}

\rev{These 36 state--time pairs are prescribed rather than taken from SFS trajectories. Two independent RQMC reference estimates, each with 65536 points per evaluation, differed by $0.01476$ on average; this indicates the numerical uncertainty of the reference but is not a rigorous error bound. MC and RQMC used $m$ from 32 to 2048, with 24 repetitions per setting. Write $(A_{\mathrm{ref}},C_{\mathrm{ref}},b_{\mathrm{ref}})$ for the reference values of the numerator, denominator, and drift, and $(\widehat A,\widehat C,\widehat b)$ for the corresponding estimates in \eqref{eq:mc_local_problem}. In each repetition, the RMSEs are root mean squares over the 36 evaluations of}
\begin{revision}
\begin{equation}
 \frac{\|\widehat A-A_{\mathrm{ref}}\|}{C_{\mathrm{ref}}},
 \qquad
 \left|\frac{\widehat C}{C_{\mathrm{ref}}}-1\right|,
 \qquad
 \|\widehat b-b_{\mathrm{ref}}\|.
 \label{eq:mg1_local_error_definitions}
\end{equation}
\end{revision}

\rev{The exponents $\gamma$ in Table~\ref{tab:mg1_local} are obtained by fitting $\mathrm{error}(m)\approx c\,m^{-\gamma}$ over the tested sizes; they are empirical, not asymptotic, rates.}

\begin{table}[!tbp]
\centering
\small
\caption{\rev{Empirical error-decay exponents and errors at $m=2048$ for M/G/1 drift estimation, with errors defined by \eqref{eq:mg1_local_error_definitions}. Exponents are the fitted decay rates $\gamma>0$}}
\label{tab:mg1_local}
\begin{tabular}{lrrrr}
\toprule
Quantity&MC exponent&RQMC exponent&MC error&RQMC error\\
\midrule
Numerator&0.503&0.588&0.0791&0.0400\\
Denominator&0.501&0.596&0.0576&0.0266\\
Drift ratio&0.488&0.549&0.1878&0.0988\\
\bottomrule
\end{tabular}
\end{table}

\rev{RQMC gave larger exponents for all three errors and reduced the drift error by $47.4\%$ at $m=2048$. For both methods, the exponent for the ratio was smaller than those for the numerator and denominator. Identity~\eqref{eq:ratio_perturbation} shows that the ratio error depends jointly on both integration errors and on the random denominator, but it does not predict this difference in fitted exponents. Comparisons at finitely many states establish neither the uniform moment conditions \eqref{eq:local_ratio_moments} nor uniform error control along trajectories with sample reuse.}

\begin{figure}[!tbp]
\centering
\includegraphics[width=.96\textwidth]{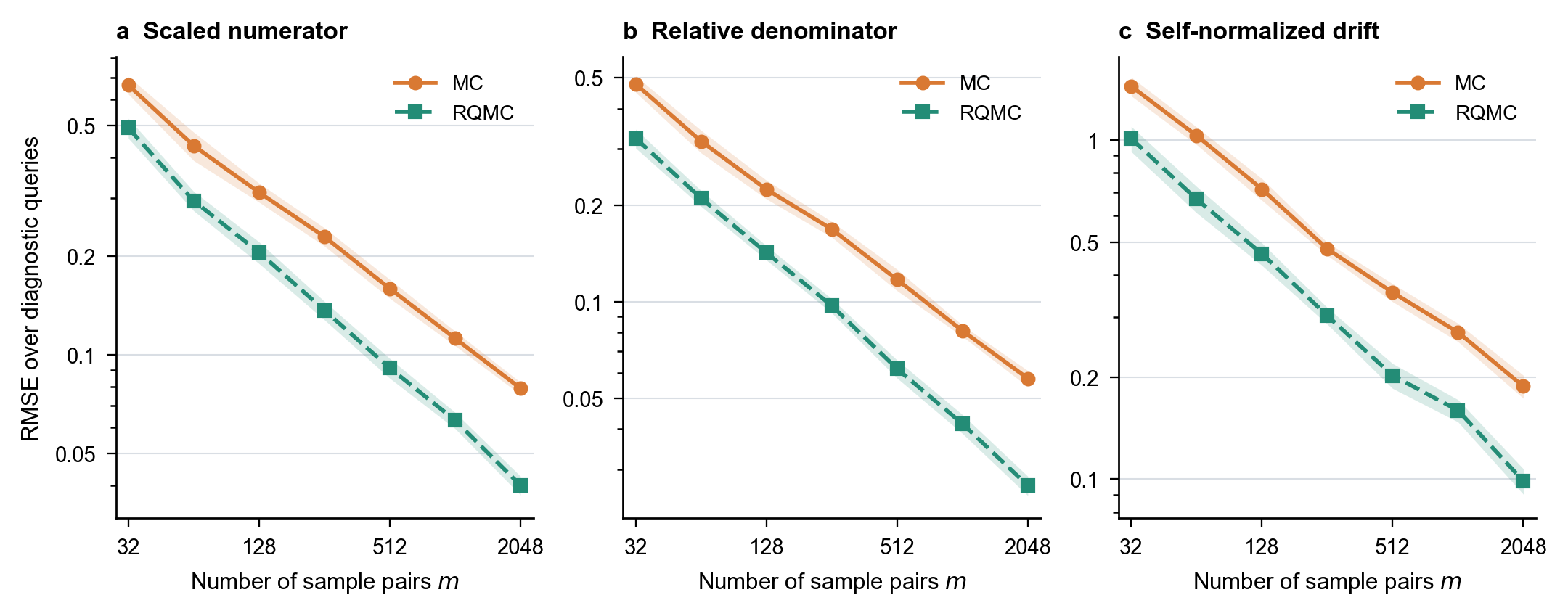}
\caption{M/G/1 drift-estimation errors versus the sample size $m$, using \eqref{eq:mg1_local_error_definitions} over the 36 prescribed evaluations. Panels show (a) numerator error scaled by the reference denominator, (b) relative denominator error, and (c) self-normalized drift error. Curves are means across 24 independent repetitions; shaded bands have half-width $1.96$ times the standard error of those means. Vertical axes are logarithmic.}
\label{fig:mg1_local}
\end{figure}

Fig.~\ref{fig:mg1_local} shows the three errors over all tested sample sizes $m$, from which the exponents in Table~\ref{tab:mg1_local} are fitted.

\rev{The terminal comparisons used RQMC samples with $M\in\{512,1024,2048,4096\}$ for all three proposals, together with independent MC sample sets for $q_2$ and $q_3$ at $M=4096$. Tables~\ref{tab:mg1_moments} and~\ref{tab:mg1_terminal} report the posterior moments of $\theta$ and the errors of the terminal distribution.}

\begin{table}[!tbp]
\centering
\small
\caption{M/G/1 posterior means and standard deviations
of $\theta$ at $M=4096$.
Standard deviations are in parentheses}
\label{tab:mg1_moments}
\begin{tabular}{lrrrr}
\toprule
Parameter&Truth&Reference&$q_2$--RQMC&$q_3$--RQMC\\
\midrule
$\theta_1$&1.0&1.0466 (0.3782)&1.0503 (0.4154)&1.0533 (0.4146)\\
$\theta_2$&4.0&4.7620 (2.5268)&4.7565 (2.5424)&4.7268 (2.5529)\\
$\theta_3$&0.2&0.2096 (0.0585)&0.2098 (0.0593)&0.2091 (0.0596)\\
\bottomrule
\end{tabular}
\end{table}

\begin{table}[!tbp]
\centering
\small
\caption{M/G/1 errors of the terminal distribution at $M=4096$.
\rev{Both distances are computed for $D_{\mathrm{ref}}^{-1}(\theta-\mu_{\mathrm{ref}}^\theta)$.}
Sliced $W_2$ is reported as mean $\pm$ standard deviation
across repetitions; \rev{local ESS is averaged at the final drift evaluation.}
\rev{Calls include the preliminary samples} but exclude bandwidth calibration}
\label{tab:mg1_terminal}
\begin{tabular}{lrrrr}
\toprule
Method&Sliced $W_2$&$W_{2,G}$&\rev{Local ESS}&Calls\\
\midrule
$q_1$--RQMC&0.1358$\pm$0.0241&0.1794&4.99&4096\\
$q_2$--MC&0.1339$\pm$0.0207&0.1574&38.26&20480\\
$q_2$--RQMC&0.1321$\pm$0.0239&0.1603&37.50&20480\\
$q_3$--MC&0.1458$\pm$0.0612&0.1939&36.42&20480\\
$q_3$--RQMC&0.1348$\pm$0.0293&0.1634&36.04&20480\\
\bottomrule
\end{tabular}
\end{table}

\rev{At $M=4096$, RQMC with $q_2$ and $q_3$ increased the local ESS \eqref{eq:qbank_ess} at the last drift evaluation about $7.5$- and $7.2$-fold relative to $q_1$, yet $q_2$ reduced the sliced $W_2$ error by only about $3\%$, compared with $23.4\%$ at $M=512$. More balanced local weights thus did not yield a proportional terminal improvement. Local ESS concerns the conditional weights, not the overall terminal importance weights or the number of independent outputs (Remark~\ref{rem:reuse_ess}).}

\rev{Two-sided paired $t$-tests on the sliced-$W_2$ errors over the 12 matched RQMC and MC repetitions gave $p=0.823$ for $q_2$ and $p=0.609$ for $q_3$; no terminal difference was detected, although this does not establish equivalence. Thus RQMC improved drift estimation but showed no terminal advantage here, where the error of the numerical reference, the time-discretization error, the Monte Carlo error due to finitely many outputs, and the error from sharing one sample set $\mathcal B_M$ among all trajectories also contribute.}

\rev{For the fitted proposals, the call counts include the $M_0=16384$ preliminary samples. Adding the 65536 bandwidth-calibration calls gives 69632 calls for $q_1$ and 86016 for $q_2$ or $q_3$ at $M=4096$, excluding the reference computation.}

\begin{figure}[!tbp]
\centering
\includegraphics[width=.72\textwidth]{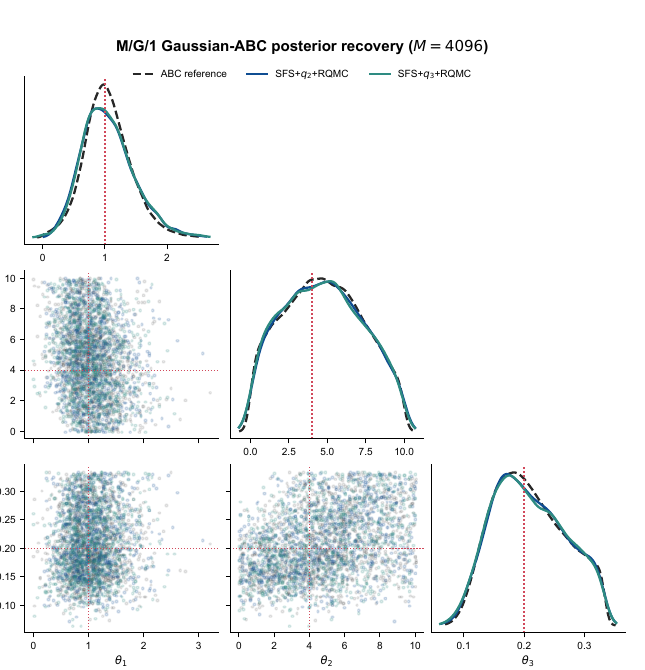}
\caption{M/G/1 marginal densities and pairwise projections
at $M=4096$. The numerical ABC reference is compared
with SFS output using sample reuse with proposals $q_2$ and $q_3$.}
\label{fig:mg1_posterior}
\end{figure}

\begin{figure}[!tbp]
\centering
\includegraphics[width=.94\textwidth]{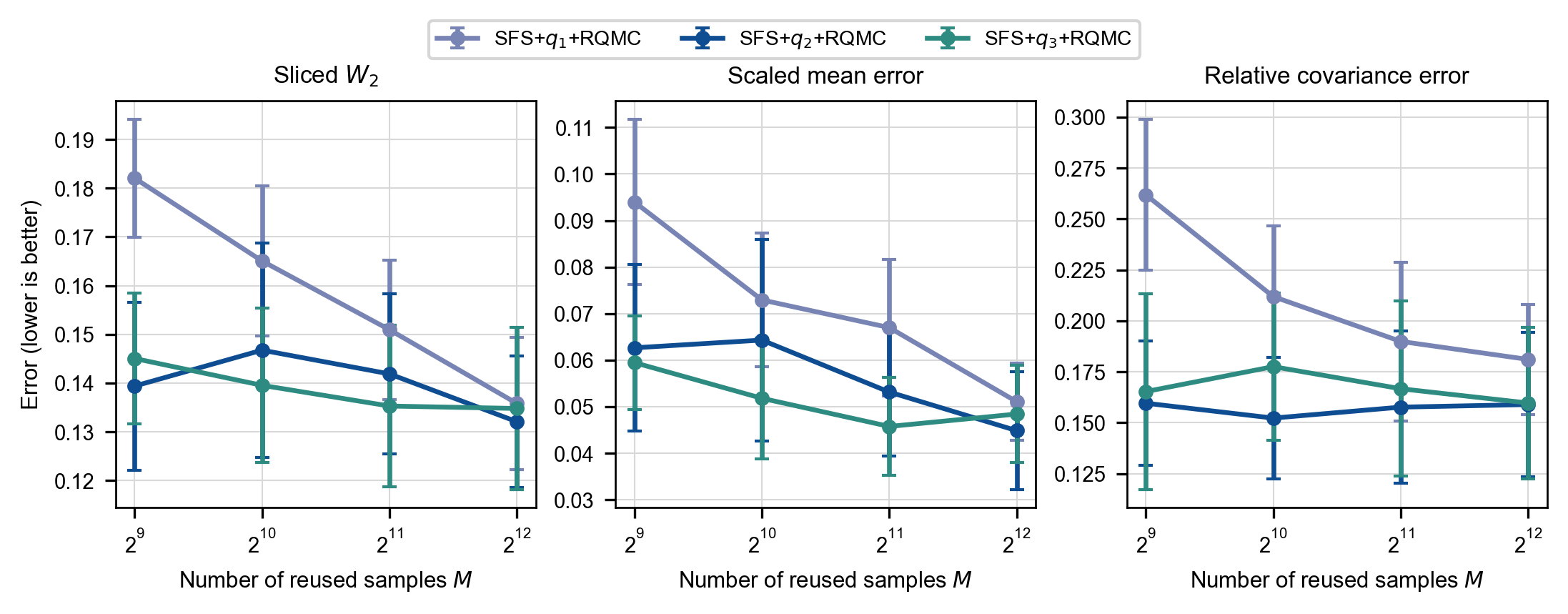}
\caption{M/G/1 terminal errors versus the number of reused samples.
Sliced $W_2$ is computed for $D_{\mathrm{ref}}^{-1}(\theta-\mu_{\mathrm{ref}}^\theta)$;
mean and covariance errors are defined in \eqref{eq:mg1_mean_error}.
Error bars are $1.96$ standard errors of the mean over
12 repetitions.
Light purple, dark blue, and green curves correspond
to $q_1$, $q_2$, and $q_3$, respectively.}
\label{fig:mg1_bank}
\end{figure}

Fig.~\ref{fig:mg1_posterior} compares the marginal densities and pairwise projections of the SFS output at $M=4096$ with the numerical reference, and Fig.~\ref{fig:mg1_bank} shows how the terminal errors change with the number $M$ of reused samples.

\subsection{\texorpdfstring{\rev{Ohio random-intercept logistic model}}{Ohio random-intercept logistic model}}

\label{sec:exp3}

\rev{The \texttt{geepack::ohio} data contain binary wheezing responses for 537 children at four visits each, ages 7--10, giving 2148 records \cite{halekoh2006geepack,fitzmaurice1993ohio}. This example uses a randomized likelihood estimator without ABC smoothing. Ages were centered at 9, so $A_{ij}\in\{-2,-1,0,1\}$; the maternal-smoking and wheezing proportions are $0.34823$ and $0.15177$. For child $i$ and visit $j$, we use the random-intercept logistic model, a generalized linear mixed model (GLMM),}
\begin{equation}
\begin{aligned}
 Y_{ij}\mid\beta,\alpha_i
 &\sim\mathrm{Bernoulli}(p_{ij}),\\
 \operatorname{logit}(p_{ij})
 &=\beta_1+\beta_2A_{ij}+\beta_3S_i+\alpha_i,\\
 \alpha_i\mid\tau
 &\sim\mathcal N(0,\tau^2),\\
 \beta&\sim\mathcal N(0,50I_3),
 \qquad
 \tau\sim\mathrm{Gamma}(1,0.1).
\end{aligned}
\label{eq:glmm_model}
\end{equation}

\rev{Responses are conditionally independent given the parameters and random effects, and the random effects are independent across children. Here $S_i$ is the maternal-smoking indicator (not a summary map), and $\beta_1$, $\beta_2$, and $\beta_3$ are the intercept, age effect, and smoking effect. The Gamma prior is parameterized by shape and rate and is placed on the standard deviation $\tau$, not on $\tau^2$. SFS runs in $x\in\R^4$, with}
\begin{equation}
 \beta_j=\sqrt{50}x_j
 \quad(j=1,2,3),
 \qquad
 \tau=-10\log\{1-\Phi_{\mathrm N}(x_4)\}.
 \label{eq:glmm_whiten}
\end{equation}

\rev{Since $\mathrm{Gamma}(1,0.1)$ is the exponential distribution with rate $0.1$, the second map in \eqref{eq:glmm_whiten} gives the stated prior on $\tau$. In this section, $\tau$ denotes the random-effect scale, not the remaining diffusion time $1-t$. The 537 random intercepts are integrated out, so the state dimension is $d=4$.}

\rev{For fixed $x$, with $\beta$ and $\tau$ defined by \eqref{eq:glmm_whiten}, set}
\begin{equation}
\begin{aligned}
 \eta_{ij}(a)
 &=\beta_1+\beta_2A_{ij}+\beta_3S_i+a,\\
 p_{ij}(a)
 &=\{1+\exp[-\eta_{ij}(a)]\}^{-1},\\
 g_i(a;x)
 &=\exp\left\{
 \sum_{j=1}^{4}
 [Y_{ij}\eta_{ij}(a)-\log(1+e^{\eta_{ij}(a)})]
 \right\}.
\end{aligned}
\label{eq:glmm_group_terms}
\end{equation}

\rev{Integrating out each random intercept gives}
\begin{equation}
 L_i(x)
 =
 \int_{\mathbb R}
 g_i(a;x)\varphi_{\tau^2}(a)\,\mathrm da,
 \qquad
 \rev{L}(x)=\prod_{i=1}^{537}L_i(x),
 \label{eq:glmm_marginal_likelihood}
\end{equation}
\rev{where $\varphi_{\tau^2}$ is the $\mathcal N(0,\tau^2)$ density. To place importance samples near the mode of each integrand, we write its logarithm, up to an additive constant independent of $a$, as}
\begin{equation}
 \ell_i(a)
 =
 \sum_{j=1}^{4}
 [Y_{ij}\eta_{ij}(a)-\log(1+e^{\eta_{ij}(a)})]
 -\frac{a^2}{2\tau^2}.
 \label{eq:glmm_log_integrand}
\end{equation}

\rev{Since $\partial_a\log(1+e^{\eta_{ij}(a)})=p_{ij}(a)$ and $\partial_ap_{ij}(a)=p_{ij}(a)\{1-p_{ij}(a)\}$,}
\begin{equation}
 \begin{aligned}
 \ell_i'(a)
 &=\sum_{j=1}^{4}[Y_{ij}-p_{ij}(a)]-a/\tau^2,\\
 -\ell_i''(a)
 &=\sum_{j=1}^{4}p_{ij}(a)[1-p_{ij}(a)]
   +\tau^{-2}>0.
 \end{aligned}
 \label{eq:glmm_log_derivatives}
\end{equation}

\rev{Since $\ell_i$ is strictly concave, its mode can be computed by Newton's method, started at zero with iterates clipped to $[-5,5]$. With the resulting estimate $\widehat a_i$ and $c_i=-\ell_i''(\widehat a_i)>0$, we use the proposal}
\begin{equation}
 \rev{q_{\mathrm{RE},i}(a\mid x)}
 =
 \mathcal N(a;\widehat a_i,1.2^2/c_i).
 \label{eq:glmm_random_effect_proposal}
\end{equation}

\rev{This one-dimensional proposal for the random effect is distinct from the parameter proposal $q$. Its location and scale follow the Laplace principle \cite{tierney1986laplace}, with the standard deviation inflated by the factor $1.2$; it serves only as an importance proposal, not as a Laplace approximation of the likelihood.}

\rev{With $J=16$, let $\alpha_{ir}=\widehat a_i+1.2c_i^{-1/2}\varepsilon_{ir}$, where, ideally, the inputs $\varepsilon_{ir}$ are independent standard normal variables, and define}
\begin{equation}
 \begin{aligned}
 \widehat L_i(x)
 &=
 \frac1J\sum_{r=1}^{J}
 \frac{g_i(\alpha_{ir};x)\varphi_{\tau^2}(\alpha_{ir})}
 {\rev{q_{\mathrm{RE},i}(\alpha_{ir}\mid x)}},\\
 \rev{\psi}(x,\omega)&=\prod_{i=1}^{537}\widehat L_i(x).
 \end{aligned}
 \label{eq:glmm_unbiased}
\end{equation}

\rev{The vector $\omega$ collects the underlying uniform inputs, as in Section~\ref{sec:exp2}. Since the Gaussian proposal has full support, the importance-sampling identity \cite[Chapter~9]{owen2013mc} gives}
\begin{align}
 \mathbb E[\widehat L_i(x)\mid x]
 &=
 \int_{\mathbb R}
 \frac{g_i(a;x)\varphi_{\tau^2}(a)}
 {\rev{q_{\mathrm{RE},i}(a\mid x)}}
 \rev{q_{\mathrm{RE},i}(a\mid x)}\,\mathrm da
 \notag\\
 &=L_i(x).
 \label{eq:glmm_importance_identity}
\end{align}

\rev{Since the randomizations are conditionally independent across children,}
\begin{equation}
 \mathbb E[\rev{\psi}(x,\omega)\mid x]
 =
 \prod_{i=1}^{537}\mathbb E[\widehat L_i(x)\mid x]
 =
 \rev{L}(x).
 \label{eq:glmm_unbiased_likelihood}
\end{equation}

\rev{Thus the ideal weight is nonnegative and unbiased even if the mode is only approximate, provided the same proposal is used for sampling and for evaluating the weight; the accuracy of the mode affects efficiency and higher moments, but not this identity. Unbiasedness does not extend to $\log\rev{\psi}$ or to the self-normalized drift, and it gives no mean-square bound for the drift. The sums and products are evaluated with log-sum-exp arithmetic for numerical stability.}

\rev{In fact, the ideal unclipped weight has infinite second moment on an open set of parameters. At $x=0$, we have $\beta=0$ and $\tau=10\log2$. For a child with two positive outcomes, $\widehat a_i=0$, the proposal variance is $s_i^2=1.2^2/(1+\tau^{-2})$, and, as $|a|\to\infty$,}
\begin{revision}
\[
 \log\frac{g_i(a;x)^2\varphi_{\tau^2}(a)^2}
 {\rev{q_{\mathrm{RE},i}(a\mid x)}}
 =\left(\frac1{2s_i^2}-\frac1{\tau^2}\right)a^2
 -4|a|+O(1).
\]
\end{revision}

\noindent\rev{The quadratic coefficient is positive, such children occur in the data, and the strict inequality persists in a neighborhood of $x=0$. Since $q_2$ and $q_3$ have full support, condition \eqref{eq:qbank_centered_moment} of Theorem~\ref{thm:qbank_pointwise} fails for the ideal weights, although the first-moment identity \eqref{eq:glmm_unbiased_likelihood} still holds.}

\rev{Each weight uses $537\times16=8592$ Gaussian inputs in addition to the parameter coordinates. The parameter and auxiliary inputs are generated from scrambled Sobol points with separate random seeds. Ideally, the auxiliary coordinates of each point are uniform on the unit cube and independent of its parameter coordinates, so that the inputs for different children are independent. Dependence across points preserves the expectation identity but violates the iid assumption of Theorem~\ref{thm:qbank_pointwise}. Before the inverse-normal transform, the uniforms are clipped to $[10^{-12},1-10^{-12}]$, which introduces an unquantified deviation from the ideal unbiased estimator. For this model, we establish neither an RQMC rate in this high dimension nor Assumption~\ref{ass:density_ratio_curvature} or the moment conditions \eqref{eq:local_ratio_moments}.}

\rev{The numerical reference combines 128-point Gauss--Hermite quadrature for each child's integral \cite{golub1969gh} with four pCN chains \cite{cotter2013pcn}, each run for 20000 steps, with the first 4000 discarded and every fourth subsequent state retained, giving 16000 samples in total. Acceptance rates were $0.235$--$0.261$, split-$\widehat R$ ranged from $1.0011$ to $1.0039$, and bulk ESS from 627 to 6949. These chain diagnostics do not bound the quadrature error, so the reference is not exact.}

\rev{The density $q_p$ for preliminary sampling was constructed from three successive sets of 4096 points, after which 32768 preliminary points were drawn from $q_p$; fitting thus cost $3\times4096+32768=45056$ weight evaluations, and the preliminary sample had ESS $985.19$. For each of $q_2$ and $q_3$, three independent repetitions used $M=16384$ pairs and $N=512$ outputs each. The proposal $q_1$ was excluded because preliminary checks showed that it covered the posterior poorly. The fitted variances were bounded below by $10^{-6}$ after the inflation in \eqref{eq:q123}.}

\rev{The time grid \eqref{eq:time_grid} was refined near its endpoint:}
\begin{equation}
 t_k=T\left[1-\left(1-\frac{k}{K}\right)^4\right],
 \qquad
 k=0,\ldots,K,
 \qquad
 K=80,\quad T=0.999999.
 \label{eq:glmm_time_grid}
\end{equation}

\rev{The output is taken at $T<1$, and the last step is}
\begin{equation}
 \Delta t_{K-1}
 =
 t_K-t_{K-1}
 =
 \frac{T}{K^4}
 \approx2.44\times10^{-8}.
 \label{eq:glmm_last_step}
\end{equation}

\rev{By \eqref{eq:euler_local}, the noise in the last step has covariance $\Delta t_{K-1}I_4$ in Gaussian coordinates. Under \eqref{eq:glmm_whiten}, its standard deviation in each $\beta_j$, which is linear in $x_j$, is}
\begin{equation}
 \sqrt{50\Delta t_{K-1}}
 \approx0.00110,
 \label{eq:glmm_last_noise}
\end{equation}
\rev{well below the reference posterior standard deviation $0.0671$ of $\beta_2$; a uniform grid with the same $K$ and $T$ would give $\sqrt{50T/K}\approx0.791$. The calculation does not apply directly to $\tau$, which depends nonlinearly on $x_4$. Small final-step noise does not by itself guarantee accuracy along the whole trajectory; a guarantee for this configuration would require error control for nonuniform grids, a time cutoff, and sample reuse, which Theorem~\ref{thm:path_error} (uniform grid, terminal time one) does not provide.}

\rev{The diagnostics use $\theta=(\beta_1,\beta_2,\beta_3,\tau^2)^\top$, that is, they report the random-effect variance, although the prior and the transformation are defined for $\tau$. For the numerical reference moments $(\mu_{\mathrm{ref}}^\theta,\Sigma_{\mathrm{ref}}^\theta)$, define}
\begin{equation}
 D_{\mathrm{ref}}=\operatorname{diag}
 \bigl(\sqrt{(\Sigma_{\mathrm{ref}}^\theta)_{11}},\ldots,
       \sqrt{(\Sigma_{\mathrm{ref}}^\theta)_{44}}\bigr).
 \label{eq:glmm_reference_scale}
\end{equation}

\rev{For the output moments $(\widehat\mu_{\mathrm{out}}^\theta,\widehat\Sigma_{\mathrm{out}}^\theta)$, set $\widehat\Sigma_{\mathrm{out},s}=D_{\mathrm{ref}}^{-1}\widehat\Sigma_{\mathrm{out}}^\theta D_{\mathrm{ref}}^{-1}$ and $\Sigma_{\mathrm{ref},s}=D_{\mathrm{ref}}^{-1}\Sigma_{\mathrm{ref}}^\theta D_{\mathrm{ref}}^{-1}$. The errors are}
\begin{equation}
 E_{\mu,\theta}
 =
 \frac{\|D_{\mathrm{ref}}^{-1}
 (\widehat\mu_{\mathrm{out}}^\theta
 -\mu_{\mathrm{ref}}^\theta)\|}{\sqrt d},
 \qquad
 E_{\Sigma,\theta}
 =
 \frac{\|\widehat\Sigma_{\mathrm{out},s}
 -\Sigma_{\mathrm{ref},s}\|_F}
 {\|\Sigma_{\mathrm{ref},s}\|_F}.
 \label{eq:glmm_standardized_errors}
\end{equation}

\rev{Here $d=4$. The diagnostic $W_{2,G}^{\mathrm{std}}$ is \eqref{eq:gaussian_w2} computed after the same reference centering and coordinatewise standardization; its scale therefore differs from that in Section~\ref{sec:exp1}, and it is not the full $W_2$ distance between the joint distributions.}

Table~\ref{tab:glmm_errors} reports the error diagnostics and the local ESS at the last drift evaluation for $q_2$ and $q_3$, and Table~\ref{tab:glmm_moments} compares the resulting posterior means and standard deviations with the reference.

\begin{table}[!tbp]
\centering
\small
\caption{Ohio GLMM posterior diagnostics.
\rev{The three error measures are reported as means $\pm$ standard deviations across three independent sets of samples used for drift estimation. Local ESS \eqref{eq:qbank_ess} measures weight concentration at the last drift evaluation; it is distinct from the overall terminal importance-weight ESS in Remark~\ref{rem:reuse_ess}}}
\label{tab:glmm_errors}
\begin{tabular}{lrrrr}
\toprule
Method&$E_{\mu,\theta}$&$E_{\Sigma,\theta}$
&$W_{2,G}^{\mathrm{std}}$&\rev{Local ESS}\\
\midrule
$q_2$--RQMC&0.105$\pm$0.054&0.226$\pm$0.118&0.315$\pm$0.128&1.025\\
$q_3$--RQMC&0.081$\pm$0.024&0.154$\pm$0.016&0.232$\pm$0.039&1.020\\
\bottomrule
\end{tabular}
\end{table}

\begin{table}[!tbp]
\centering
\small
\caption{Ohio GLMM posterior means (standard deviations)
and the ratio of the $q_3$ standard deviation to the
reference value.
\rev{For $q_2$ and $q_3$, each entry averages the means and sample standard deviations (divisor $N-1$) computed separately in the three runs. The SD ratio is the average of the three ratios of the $q_3$ standard deviation to the reference value}}
\label{tab:glmm_moments}
\begin{tabular}{lrrrr}
\toprule
Parameter&Reference&$q_2$--RQMC&$q_3$--RQMC&SD ratio\\
\midrule
$\beta_1$&$-3.128$ (0.230)&$-3.145$ (0.226)&$-3.143$ (0.235)&1.024\\
$\beta_2$&$-0.175$ (0.067)&$-0.184$ (0.075)&$-0.181$ (0.073)&1.089\\
$\beta_3$&$0.399$ (0.283)&$0.397$ (0.286)&$0.415$ (0.288)&1.020\\
$\tau^2$&$4.877$ (0.867)&$4.966$ (0.911)&$4.918$ (0.869)&1.003\\
\bottomrule
\end{tabular}
\end{table}

\rev{The mixture $q_3$ gave smaller values of all three error diagnostics and less variation across sample sets; its marginal standard-deviation ratios ranged from $1.003$ to $1.089$, with the largest deviation for $\beta_2$. With only three repetitions per proposal, these comparisons are descriptive and do not show that $q_3$ is better in general. The local ESS \eqref{eq:qbank_ess} at the last drift evaluation was about $1.02$ for both proposals; by Remark~\ref{rem:reuse_ess}, this occurs even under exact sampling of the fixed-sample terminal measure and does not by itself indicate posterior error or instability. Finally, agreement of moments and one-dimensional marginals does not establish accuracy of the joint distribution.}

\begin{figure}[!tbp]
\centering
\includegraphics[width=.94\textwidth]{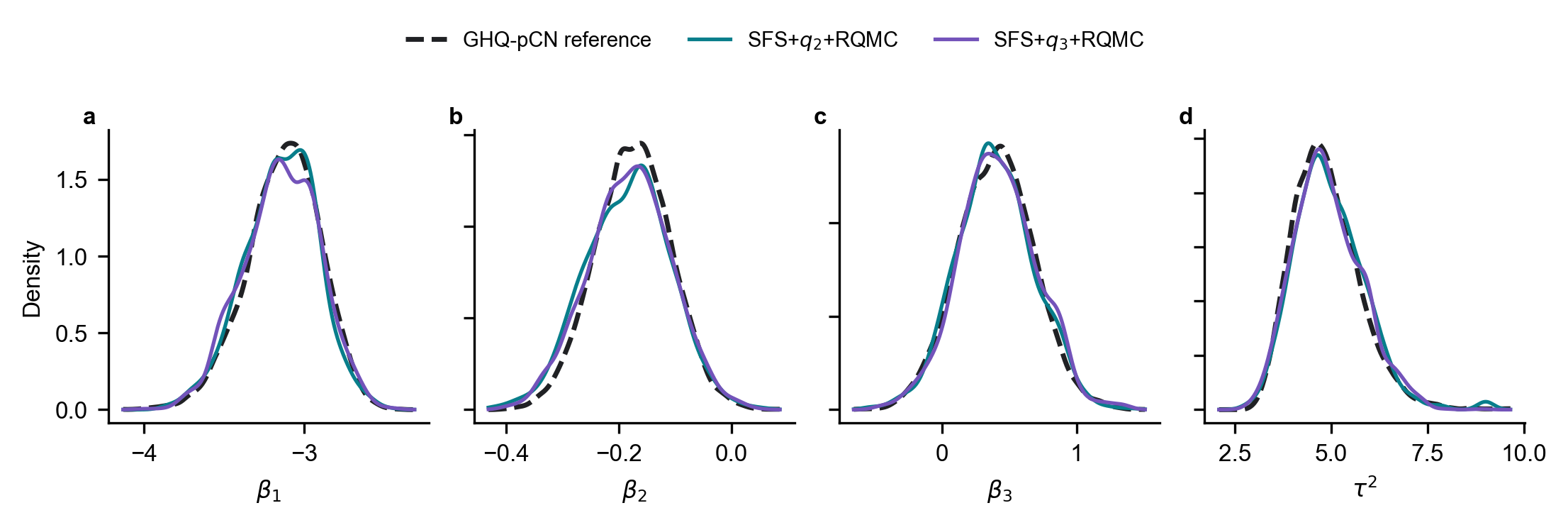}
\caption{Ohio GLMM posterior marginals.
For each of $q_2$ and $q_3$, the $3\times512=1536$ outputs of the three runs are pooled into one kernel density estimate. The reference density, computed by Gauss--Hermite quadrature and pCN, uses the 16000 retained samples.
Panels (a)--(d) show $\beta_1$, $\beta_2$, $\beta_3$,
and $\tau^2$, respectively.}
\label{fig:glmm_marginals}
\end{figure}

Fig.~\ref{fig:glmm_marginals} compares the marginal posterior densities of $\beta_1$, $\beta_2$, $\beta_3$, and $\tau^2$ obtained with $q_2$ and $q_3$ with the reference computed by Gauss--Hermite quadrature and pCN.

\rev{Proposal fitting and the six sample sets required $45056+6\times16384=143360$ evaluations of $\psi$, each comprising $537\times16$ random-effect importance terms. These counts are not directly comparable with the M/G/1 simulator calls and do not measure wall-clock time; the reference computation and the state-dependent weight calculations are additional costs.}

\section{Conclusion}
\label{sec:conclusion}
\label{sec:discussion}

We developed a likelihood-free Schr\"odinger--F\"ollmer sampler whose drift is estimated jointly from Gaussian samples and simulator outputs, requiring neither pointwise likelihood values nor simulator derivatives. Importance sampling with a fixed proposal allows a single set of simulator outputs to be reused across time steps and trajectories; this reduces the number of simulator calls, although the state-dependent weights must still be recomputed.

For likelihood-free SFS, we gave sufficient curvature conditions for ABC targets, established drift regularity, and derived a terminal error bound that separates integration and discretization errors. For sample reuse with iid samples, we proved pointwise consistency and asymptotic normality of the drift estimator, which yield a variance-based proposal criterion, and terminal consistency under a finite posterior second moment; unconditional $W_2$ convergence requires additional moments of the proposal. Some of these assumptions are restrictive: the Gaussian benchmark violates the state-uniform mean-square assumption of Theorem~\ref{thm:path_error}, and the ideal Ohio weights violate the second-moment assumption of the drift CLT. Neither assumption is needed for terminal consistency under sample reuse.

In the experiments, sample reuse greatly reduced the number of simulator calls, and RQMC improved drift estimation. Fitted proposals improved centering and local weight balance, but these gains did not always translate into proportional terminal improvements: the concentrated Gaussian example is limited by an Euler covariance floor, and the M/G/1 example showed no terminal advantage of RQMC. The Ohio example applies the construction to an unbiased randomized likelihood without ABC. Quantitative terminal bounds in the sample size and time step under sample reuse, covering the RQMC rules, clipped inputs, and time cutoff used in the experiments, remain to be developed.

\appendices
\section{Proofs for likelihood-free SFS}
\label{sec:appendix_fresh}

Throughout this appendix, Assumption~\ref{ass:density_ratio_curvature} holds and we write
\[
 \tau=1-t,\qquad L_V=\max\{\lambda_+,\kappa\},\qquad L_b=\frac{L_V}{1-\kappa}.
\]
In particular, $\|\nabla^2V(z)\|_{\mathrm{op}}\le L_V$ for all $z\in\R^d$.

\subsection{Proof of Theorem~\ref{thm:path_error}}
\label{sec:proof_fresh}

The proof is based on a synchronous coupling of the exact diffusion and the Euler scheme. We first establish the regularity of the exact drift in Lemma~\ref{lem:drift_regularity} and bound the drift-estimation error at a fixed state in Lemma~\ref{lem:local_mse}; the two lemmas are then combined through a discrete Gronwall argument.
For $(x,t)\in\R^d\times[0,1)$, define the probability measure $\mu_{x,t}$ and its potential $\mathcal U_{x,t}$ by
\begin{equation}
 \mu_{x,t}(\mathrm dz)=\frac{f(z)\varphi_\tau(z-x)}{Q_\tau f(x)}\,\mathrm dz,
 \qquad \mathcal U_{x,t}(z)=V(z)+\frac{\|z-x\|^2}{2\tau}.
 \label{eq:bridge_tilt}
\end{equation}
Since $f=e^{-V}$, the function $\mathcal U_{x,t}$ is the negative log-density of $\mu_{x,t}$ up to an additive constant.

\begin{lemma}
\label{lem:drift_regularity}
The drift $b$ extends continuously to $t=1$ by $b(x,1)=-\nabla V(x)$, the equation \eqref{eq:sfs_sde} has a unique strong solution on $[0,1]$, and $X_1\sim\pi_{\ABC}$.
Moreover, there are constants $B_*$ and $C_X$, depending only on $B_0$, $\lambda_+$, and $\kappa$, such that for all $x,x'\in\R^d$ and $0\le s\le t\le1$,
\begin{align}
 \|b(x,t)-b(x',t)\|&\le L_b\|x-x'\|,
 \label{eq:drift_lipschitz}\\
 \|b(x,t)\|&\le B_*\sqrt d+L_b\|x\|,
 \label{eq:drift_growth}\\
 \sup_{0\le u\le1}\E\|X_u\|^2&\le C_Xd,
 \label{eq:path_moment}\\
 \E\|b(X_t,t)-b(X_s,s)\|^2&\le dL_b^2(t-s).
 \label{eq:drift_time}
\end{align}
\end{lemma}

\begin{proof}
We first prove \eqref{eq:drift_lipschitz} and \eqref{eq:drift_growth} for $t<1$.
By Assumption~\ref{ass:density_ratio_curvature},
$\nabla_z^2\mathcal U_{x,t}\succeq(\tau^{-1}-\kappa)I_d$ with $\tau^{-1}-\kappa\ge1-\kappa>0$, so $\mu_{x,t}$ is strongly log-concave.
By the Brascamp--Lieb variance inequality, its Poincar\'e constant satisfies
\begin{equation}
 C_P(x,t)\le(\tau^{-1}-\kappa)^{-1}=\frac{\tau}{1-\kappa\tau}.
 \label{eq:poincare}
\end{equation}
By Taylor's formula, $V(z)\ge V(0)+\nabla V(0)^\top z-\kappa\|z\|^2/2$.
Since $\tau^{-1}>\kappa$, the function $f(z)\varphi_\tau(z-x)$ and its derivatives in $x$ are dominated by Gaussian functions of $z$, locally uniformly in $x$.
Hence we may differentiate under the integral sign and integrate by parts.
Differentiating $Q_\tau f$ gives $b(x,t)=\E_{\mu_{x,t}}[(Z-x)/\tau]$, and integrating $\nabla_ze^{-\mathcal U_{x,t}}$ over $\R^d$ gives $\E_{\mu_{x,t}}\nabla\mathcal U_{x,t}(Z)=0$.
It follows that
\begin{equation}
 b(x,t)=\E_{\mu_{x,t}}\frac{Z-x}{\tau}
 =-\E_{\mu_{x,t}}\nabla V(Z).
 \label{eq:drift_potential_mean}
\end{equation}

Differentiating the density of $\mu_{x,t}$ with respect to $x$ in \eqref{eq:drift_potential_mean} gives
\[
 \nabla_xb(x,t)=-\tau^{-1}\operatorname{Cov}_{\mu_{x,t}}(\nabla V(Z),Z).
\]
For unit vectors $u,v\in\R^d$, the Cauchy--Schwarz inequality and \eqref{eq:poincare} give
\begin{equation}
 |u^\top\nabla_xb(x,t)v|
 \le\frac1\tau\operatorname{Var}_{\mu_{x,t}}(u^\top\nabla V(Z))^{1/2}
 \operatorname{Var}_{\mu_{x,t}}(v^\top Z)^{1/2}
 \le\frac{C_P(x,t)L_V}{\tau}
 \le\frac{L_V}{1-\kappa\tau}\le L_b,
 \label{eq:drift_jacobian_bound}
\end{equation}
where in the second inequality we use $\operatorname{Var}(u^\top\nabla V(Z))\le C_P\E\|\nabla^2V(Z)u\|^2\le C_PL_V^2$ and $\operatorname{Var}(v^\top Z)\le C_P$.
Therefore, \eqref{eq:drift_lipschitz} holds for $t<1$.

For \eqref{eq:drift_growth}, it suffices to bound $b(0,t)$. Let $z_t^*$ be the minimizer of $\mathcal U_{0,t}$.
Since $\mathcal U_{0,t}$ is $(1-\kappa)$-strongly convex and $\nabla\mathcal U_{0,t}(0)=\nabla V(0)$, we have $\|z_t^*\|\le\|\nabla V(0)\|/(1-\kappa)\le B_0\sqrt d/(1-\kappa)$.
Integrating $\nabla_z\cdot[(z-z_t^*)e^{-\mathcal U_{0,t}(z)}]$ over $\R^d$ gives
$d=\E_{\mu_{0,t}}[(Z-z_t^*)^\top\nabla\mathcal U_{0,t}(Z)]\ge(1-\kappa)\E_{\mu_{0,t}}\|Z-z_t^*\|^2$.
Therefore,
\begin{equation}
 \E_{\mu_{0,t}}\|Z\|
 \le\left(\frac{B_0}{1-\kappa}+\frac1{\sqrt{1-\kappa}}\right)\sqrt d.
 \label{eq:bridge_origin_moment}
\end{equation}
Since $\|\nabla V(z)\|\le B_0\sqrt d+L_V\|z\|$, \eqref{eq:drift_potential_mean} and \eqref{eq:bridge_origin_moment} give $\|b(0,t)\|\le B_*\sqrt d$ with
$B_*=B_0+L_V[B_0/(1-\kappa)+(1-\kappa)^{-1/2}]$.
Together with \eqref{eq:drift_lipschitz}, this proves \eqref{eq:drift_growth} for $t<1$.

We next extend the drift to $t=1$. Let $z^*_{x,t}$ be the minimizer of $\mathcal U_{x,t}$.
The same strong convexity and integration-by-parts arguments give
\[
 \|z^*_{x,t}-x\|\le\frac{\tau\|\nabla V(x)\|}{1-\kappa\tau},
 \qquad
 \E_{\mu_{x,t}}\|Z-z^*_{x,t}\|^2\le\frac{d\tau}{1-\kappa\tau}.
\]
By \eqref{eq:drift_potential_mean} and the Lipschitz continuity of $\nabla V$,
$\|b(x,t)+\nabla V(x)\|\le L_V\E_{\mu_{x,t}}\|Z-x\|\to0$ locally uniformly in $x$ as $t\uparrow1$.
Hence $b$ extends continuously to $t=1$ by $b(x,1)=-\nabla V(x)$, and \eqref{eq:drift_lipschitz} and \eqref{eq:drift_growth} remain valid at $t=1$.
Since the drift is globally Lipschitz in $x$ with linear growth, uniformly in $t\in[0,1]$, \eqref{eq:sfs_sde} has a unique strong solution on $[0,1]$.
By It\^o's formula for $\|X_t\|^2$, \eqref{eq:drift_growth}, and Gronwall's inequality, we obtain \eqref{eq:path_moment}.

To prove \eqref{eq:drift_time}, note that for $t<1$ the function $h=Q_\tau f$ solves the backward heat equation $\partial_th+\frac12\Delta h=0$.
Hence $b=\nabla_x\log h$ satisfies $\partial_tb+\frac12\Delta b+(\nabla_xb)b=0$, and It\^o's formula gives $\mathrm db(X_t,t)=\nabla_xb(X_t,t)\,\mathrm dB_t$.
Since $\|\nabla_xb\|_F^2\le d\|\nabla_xb\|_{\mathrm{op}}^2\le dL_b^2$ by \eqref{eq:drift_jacobian_bound}, it follows from the It\^o isometry that \eqref{eq:drift_time} holds for $t<1$; the case $t=1$ follows by continuity.

It remains to identify the terminal law. Let $B^0$ be a Brownian motion under a probability measure $\mathbb P_0$, and define $\mathrm d\mathbb P_*=f(B_1^0)\,\mathrm d\mathbb P_0$, which is a probability measure because $\int f\varphi=1$.
Its density process is $M_t=Q_{1-t}f(B_t^0)$, and $\mathrm dM_t=M_tb(B_t^0,t)^\top\mathrm dB_t^0$.
By Girsanov's theorem on $[0,T]$ for each $T<1$, the process $B^0$ solves \eqref{eq:sfs_sde} under $\mathbb P_*$.
Pathwise uniqueness implies uniqueness in law, so $X$ and $B^0$ under $\mathbb P_*$ have the same law on $[0,T]$ for every $T<1$; by path continuity, $X_1$ has the law of $B^0_1$ under $\mathbb P_*$, whose density is $f\varphi=\pi_{\ABC}$.
This proves the desired result.
\end{proof}

We next bound the drift-estimation error at a fixed state.
For fixed $(x,t)$ with $t<1$, let $A(x,t)=\E[W(Z,\xi)(Z-x)]$, with expectation under \eqref{eq:joint_sampling}, let $C(x,t)$ be as in Section~\ref{sec:joint_drift}, and let $\widehat A_m$ and $\widehat C_m$ be the corresponding sample averages in \eqref{eq:mc_local_problem}.
Direct subtraction gives
\begin{equation}
 \widehat b_m-b
 =\frac{C(\widehat A_m-A)-A(\widehat C_m-C)}{(1-t)C\widehat C_m}.
 \label{eq:ratio_perturbation}
\end{equation}
Define
\begin{equation}
 \chi_{x,t}(Z,\xi)=\frac{W(Z,\xi)}{C(x,t)}
 \left\{\frac{Z-x}{1-t}-b(x,t)\right\}.
 \label{eq:influence}
\end{equation}
Since $A=(1-t)Cb$, we have $\E\chi_{x,t}(Z,\xi)=0$, and \eqref{eq:ratio_perturbation} can be rewritten as
\begin{equation}
 \widehat b_m-b
 =\frac{C}{\widehat C_m}\frac1m\sum_{i=1}^m\chi_{x,t}(Z_i,\xi_i),
 \label{eq:influence_exact}
\end{equation}
which is the vector form of the self-normalized ratio identity \cite[Section~9.2]{owen2013mc}.

\begin{lemma} 
\label{lem:local_mse}
Fix $(x,t)$ with $t<1$ and $0<C(x,t)<\infty$, let $(Z_i,\xi_i)$, $i=1,\ldots,m$, be independent with law \eqref{eq:joint_sampling}, and suppose that $\widehat C_m(x,t)>0$ almost surely.
If there are finite $D(t),R(t)\ge0$ such that
\begin{equation}
 \E\!\left[\left(\frac{C(x,t)}{\widehat C_m(x,t)}\right)^4\right]\le D(t),
 \qquad \E\|\chi_{x,t}(Z,\xi)\|^4\le R(t)d^2,
 \label{eq:local_ratio_moments}
\end{equation}
then
\begin{equation}
 \E\|\widehat b_m(x,t)-b(x,t)\|^2
 \le\frac{\rho(t)d}{m},\qquad \rho(t)=\sqrt{3D(t)R(t)}.
 \label{eq:local_mse}
\end{equation}
\end{lemma}

\begin{proof}
Write $\chi_i=\chi_{x,t}(Z_i,\xi_i)$.
Since the $\chi_i$ are independent and centered, expanding the fourth power, we have
\[
 \E\left\|\sum_{i=1}^m\chi_i\right\|^4
 =\sum_{i=1}^m\E\|\chi_i\|^4
 +\sum_{i\ne l}\left(\E\|\chi_i\|^2\E\|\chi_l\|^2+2\E(\chi_i^\top\chi_l)^2\right)
 \le3m^2R(t)d^2,
\]
where the inequality used $\E(\chi_i^\top\chi_l)^2\le\E\|\chi_i\|^2\E\|\chi_l\|^2\le R(t)d^2$ for $i\ne l$.
By \eqref{eq:influence_exact} and the Cauchy--Schwarz inequality, we obtain
\begin{equation}
 \E\|\widehat b_m-b\|^2
 \le D(t)^{1/2}\left(\E\left\|\frac1m\sum_{i=1}^m\chi_i\right\|^4\right)^{1/2}
 \le\frac{\sqrt{3D(t)R(t)}\,d}{m}.
 \label{eq:local_mse_proof}
\end{equation}
This proves the desired result.
\end{proof}

If \eqref{eq:local_ratio_moments} holds uniformly over the states visited by the scheme, then conditioning on $\mathcal F_{t_k}$ and applying Lemma~\ref{lem:local_mse} at $(x,t)=(\widetilde X_{t_k},t_k)$ gives \eqref{eq:conditional_mse}.
Based on Lemmas~\ref{lem:drift_regularity} and~\ref{lem:local_mse}, we prove Theorem~\ref{thm:path_error}.

\begin{proof}[Proof of Theorem~\ref{thm:path_error}]
We couple the exact diffusion and the scheme by using the same Brownian increments \cite{huang2021sfs}.
Let $e_k=X_{t_k}-\widetilde X_{t_k}$, so that $e_0=0$.
Subtracting the Euler update \eqref{eq:euler_local} from the integral form of \eqref{eq:sfs_sde} gives
$e_{k+1}=e_k+ha_k+\mathcal R_k+h\zeta_k$, where
\begin{equation}
 \begin{aligned}
 a_k&=b(X_{t_k},t_k)-b(\widetilde X_{t_k},t_k),\\
 \mathcal R_k&=\int_{t_k}^{t_{k+1}}[b(X_u,u)-b(X_{t_k},t_k)]\,\mathrm du,\\
 \zeta_k&=b(\widetilde X_{t_k},t_k)-\widehat b_m(\widetilde X_{t_k},t_k).
 \end{aligned}
 \label{eq:coupling_terms}
\end{equation}
It suffices to estimate each $\E\|e_{k+1}\|^2$ in terms of $\E\|e_k\|^2$. For $h\le1$, Young's inequality $2u^\top v\le h\|u\|^2+h^{-1}\|v\|^2$ yields
\[
 \|e_{k+1}\|^2
 \le(1+h)\|e_k+ha_k\|^2+\frac4h\left(\|\mathcal R_k\|^2+h^2\|\zeta_k\|^2\right).
\]
Taking expectations, we have
\[
 \begin{aligned}
 \E\|e_{k+1}\|^2
 &\le(1+h)(1+hL_b)^2\E\|e_k\|^2+\frac4h\cdot\frac{dL_b^2h^3}{2}+4h\cdot\frac{\rho(t_k)d}{m}\\
 &\le(1+c_0h)\E\|e_k\|^2
 +2dL_b^2h^2+\frac{4dh\rho(t_k)}m,
 \end{aligned}
\]
where the first inequality follows from $\|e_k+ha_k\|\le(1+hL_b)\|e_k\|$ by \eqref{eq:drift_lipschitz}, from $\E\|\mathcal R_k\|^2\le h\int_0^hdL_b^2u\,\mathrm du=dL_b^2h^3/2$ by the Cauchy--Schwarz inequality and \eqref{eq:drift_time}, and from $\E\|\zeta_k\|^2\le\rho(t_k)d/m$ by \eqref{eq:conditional_mse}, and the second inequality follows from $(1+h)(1+hL_b)^2\le1+c_0h$ with $c_0=1+4L_b+2L_b^2$.
Iterating from $e_0=0$ and using $(1+c_0h)^K\le e^{c_0}$ and $Kh=1$, we obtain
\begin{equation}
 \E\|e_K\|^2\le e^{c_0}\left(2dL_b^2h+\frac{4d\Lambda_h}m\right)
 \le C_{\mathrm{cpl}}\left(dh+\frac{d\Lambda_h}{m}\right),
 \label{eq:coupling_mse}
\end{equation}
for the constant $C_{\mathrm{cpl}}=e^{c_0}\max\{2L_b^2,4\}$ depending only on $L_b$.
Since $X_1\sim\pi_{\ABC}$ by Lemma~\ref{lem:drift_regularity}, we have $W_2^2(\operatorname{Law}(\widetilde X_1),\pi_{\ABC})\le\E\|e_K\|^2$.
The desired result follows from \eqref{eq:coupling_mse} with $C_{\mathrm{path}}=C_{\mathrm{cpl}}^{1/2}$.
\end{proof}

\subsection{Verifying the ABC curvature condition}
\label{sec:abc_verification}

We give sufficient conditions on the simulator density and the prior for Assumption~\ref{ass:density_ratio_curvature}.
The derivatives below are used only to verify this assumption; the sampler does not require them.
For fixed $y$ and $\delta>0$, write $\mathcal K(\xi)=K_\delta(S(\xi),S(y))$ and $L=L_{\ABC}>0$, and define
\begin{equation}
 s_x(\xi)=\nabla_x\log p(\xi\mid x),\qquad
 H_x(\xi)=\nabla_x^2\log p(\xi\mid x),\qquad
 \nu_{x,\delta}(\mathrm d\xi)=\frac{\mathcal K(\xi)p(\xi\mid x)}{L(x)}\,\mathrm d\xi.
 \label{eq:abc_tilt}
\end{equation}

\begin{lemma}
\label{lem:abc_curvature}
Suppose that $p(\xi\mid x)>0$ on a support independent of $x$, that $p(\xi\mid\cdot)\in C^2(\R^d)$ for almost every $\xi$, and that $\mathcal Kp$, $\mathcal K\nabla p$, $\mathcal K\nabla^2p$, and $\mathcal K\|\nabla p\|^2/p$ are dominated by integrable functions of $\xi$, locally uniformly in $x$.
Suppose also that there are constants $0\le\kappa<1$ and $\lambda_+,\Gamma_0\ge0$ such that, for every $x\in\R^d$ and $\nu=\nu_{x,\delta}$,
\begin{equation}
 \lambda_{\max}\!\left(\E_\nu H_x+\Cov_\nu(s_x)\right)\le\kappa,
 \qquad -\lambda_{\min}(\E_\nu H_x)\le\lambda_+,
 \qquad \|\E_{\nu_{0,\delta}}s_0\|\le\Gamma_0\sqrt d.
 \label{eq:simulator_curvature}
\end{equation}
If $\pi_0=\varphi$, then Assumption~\ref{ass:density_ratio_curvature} holds with $B_0=\Gamma_0$.
\end{lemma}

\begin{proof}
By the domination assumptions, we may differentiate $L(x)=\int\mathcal K(\xi)p(\xi\mid x)\,\mathrm d\xi$ twice under the integral sign, which gives
\begin{align}
 \nabla[-\log L(x)]&=-\E_{\nu_{x,\delta}}s_x(\xi),
 \label{eq:abc_gradient}\\
 \nabla^2[-\log L(x)]&=-\E_{\nu_{x,\delta}}H_x(\xi)
 -\Cov_{\nu_{x,\delta}}(s_x(\xi)).
 \label{eq:abc_hessian}
\end{align}
By the first condition in \eqref{eq:simulator_curvature}, $\nabla^2[-\log L]\succeq-\kappa I_d$.
Since covariance matrices are positive semidefinite, the second condition gives $\nabla^2[-\log L]\preceq-\E_\nu H_x\preceq\lambda_+I_d$, and the third gives $\|\nabla[-\log L](0)\|\le\Gamma_0\sqrt d$.
For a positive $C^2$ prior, $V=-\log(\pi_0L/\varphi)$ up to an additive constant, so
\begin{equation}
 \nabla^2V=\nabla^2[-\log L]+\nabla^2[-\log\pi_0]-I_d.
 \label{eq:prior_curvature}
\end{equation}
If $\pi_0=\varphi$, the last two terms cancel and $\nabla V=\nabla[-\log L]$. This proves the desired result.
\end{proof}

For other priors, the curvature of $-\log\pi_0$ enters through \eqref{eq:prior_curvature}, and the gradient at the origin is bounded by using $\nabla V(0)=-\E_{\nu_{0,\delta}}s_0-\nabla\log\pi_0(0)$.

\section{Proofs for likelihood-free SFS with sample reuse}
\label{sec:appendix_reuse}

\subsection{Proof of Theorem~\ref{thm:reuse_posterior_terminal}}
\label{sec:proof_reuse}

Throughout this subsection, the samples are fixed until the final step of the proof, and all expectations are conditional on them.
Set
\[
 H_t(x)=\sum_{i=1}^Mc_i\varphi_{1-t}(Z_i-x),\qquad 0\le t<1.
\]
If $\sum_iv_i>0$, then $H_t>0$, $\partial_tH_t=-\tfrac12\Delta H_t$, and $\widehat b_{M,q}=\nabla\log H_t$.
Consider the diffusion
\[
 \mathrm dX_t^M=\widehat b_{M,q}(X_t^M,t)\,\mathrm dt+\mathrm dB_t,
 \qquad X_0^M=0.
\]

\begin{lemma}
\label{lem:reuse_exact}
Suppose that $c_i\in[0,\infty)$ for all $i$ and $\sum_iv_i>0$.
Then $X^M$ has a unique strong solution on $[0,1)$, and its path distribution is that of $tZ_I+\beta_t$, where $\mathbb P(I=i)=a_{i,M}$ and $\beta$ is an independent standard Brownian bridge from zero to zero.
In particular, $X^M$ extends continuously to $t=1$,
\begin{equation}
 \mathcal L(X_t^M\mid\mathcal B_M)
 =\sum_{i=1}^M a_{i,M}\mathcal N(tZ_i,t(1-t)I_d),\quad 0<t<1,
 \qquad
 \mathcal L(X_1^M\mid\mathcal B_M)=\nu_M,
 \label{eq:reuse_exact_marginals}
\end{equation}
and, for $0\le T<1$,
\begin{equation}
 W_2^2\!\left(\mathcal L(X_T^M\mid\mathcal B_M),\nu_M\right)
 \le (1-T)^2\sum_{i=1}^M a_{i,M}\|Z_i\|^2+dT(1-T).
 \label{eq:reuse_time_cutoff}
\end{equation}
\end{lemma}

\begin{proof}
The proof is based on an explicit construction of the transition densities. For $0\le s<t<1$, define
\[
 p_{s,t}(x,z)=\varphi_{t-s}(z-x)\frac{H_t(z)}{H_s(x)}.
\]
By the Gaussian convolution identity $\int\varphi_{t-s}(z-x)\varphi_{1-t}(Z_i-z)\,\mathrm dz=\varphi_{1-s}(Z_i-x)$, each $p_{s,t}(x,\cdot)$ integrates to one.
Hence these transition densities define a Markov process with drift $\nabla\log H_t$, whose density at time $t$, when started at zero, is $\varphi_t(x)H_t(x)/H_0(0)$.
The identity
\[
 \varphi_t(x)\varphi_{1-t}(Z_i-x)
 =\varphi(Z_i)\varphi_{t(1-t)}(x-tZ_i)
\]
and $H_0(0)=\sum_iv_i$ give \eqref{eq:reuse_exact_marginals} for $0<t<1$.
The same transition densities arise by first choosing $I$ with probabilities $a_{i,M}$ and then running a Brownian bridge from zero to $Z_I$, which identifies the path distribution and its continuous endpoint.
On every $[0,T]$ with $T<1$, the drift is smooth, globally Lipschitz in space, and of linear growth; see \eqref{eq:reuse_drift_jacobian} below.
Existence and pathwise uniqueness therefore identify this process with the strong solution.
Finally, coupling $X_T^M=TZ_I+\beta_T$ with $Z_I$ and using the independence of $\beta_T$ and $Z_I$ together with $\mathbb E\|\beta_T\|^2=dT(1-T)$ gives \eqref{eq:reuse_time_cutoff}.
This proves the desired result.
\end{proof}

We now prove the Euler convergence \eqref{eq:reuse_euler_terminal}.
Removing the terms with $c_i=0$ and merging coincident nodes, adding their coefficients and terminal probabilities, changes neither $\widehat b_{M,q}$ nor $\nu_M$.
Without loss of generality, we therefore assume that $(Z_i,c_i,a_{i,M})_{i=1}^n$ consists of distinct nodes with positive coefficients.
Write
\[
 s=1-t,\qquad
 w_i(x,t)=\frac{c_i\varphi_s(Z_i-x)}{H_t(x)},\qquad
 m_M(x,t)=\sum_{i=1}^nw_i(x,t)Z_i,\qquad
 R=\max_i\|Z_i\|.
\]
Then $\widehat b_{M,q}(x,t)=(m_M(x,t)-x)/s$ and $\|m_M(x,t)\|\le R$.
Since $\nabla_xw_i=w_i(Z_i-m_M)/s$, we have
\begin{equation}
 \nabla_x\widehat b_{M,q}(x,t)
 =-\frac{I_d}{s}+\frac{\operatorname{Cov}_{w(x,t)}(Z)}{s^2},\qquad
 \|\nabla_x\widehat b_{M,q}(x,t)\|_{\mathrm{op}}
 \le L_\varepsilon:=\varepsilon^{-1}+R^2\varepsilon^{-2},\quad t\le1-\varepsilon,
 \label{eq:reuse_drift_jacobian}
\end{equation}
where $\operatorname{Cov}_{w(x,t)}$ denotes the covariance under the probabilities $w_i(x,t)$.

\begin{lemma}
\label{lem:reuse_euler}
Under the assumptions of Lemma~\ref{lem:reuse_exact}, let $\widetilde X^{M,h}$ be the Euler scheme with drift $\widehat b_{M,q}$ on a deterministic grid with $t_K=1$ and mesh $h$, driven by the same Brownian motion as $X^M$.
Then $\mathbb E\|\widetilde X_1^{M,h}-X_1^M\|^2\to0$ as $h\downarrow0$.
In particular, \eqref{eq:reuse_euler_terminal} holds.
\end{lemma}

\begin{proof}
Write $Y_k=\widetilde X_{t_k}^{M,h}$. Since $W_2^2(\mathcal L(Y_K\mid\mathcal B_M),\nu_M)\le\mathbb E\|Y_K-X_1^M\|^2$, it suffices to prove the first statement.
The proof is based on splitting $[0,1]$ into an initial interval $[0,T_h]$, on which the drift is Lipschitz, and a terminal interval $[T_h,1]$, on which the iterates are attracted to the endpoint $Z_I$.

We first estimate the error on the initial interval. Fix $\varepsilon\in(0,1)$, and let $T_h=t_j$ be the last grid point not exceeding $1-\varepsilon$.
For $h\le\varepsilon$, we have $\varepsilon\le1-T_h\le2\varepsilon$.
The backward heat equation for $H$ implies
\[
 \partial_t\widehat b_{M,q}
 +\tfrac12\Delta\widehat b_{M,q}
 +(\nabla_x\widehat b_{M,q})\widehat b_{M,q}=0,
\]
so It\^o's formula gives $\mathrm d\widehat b_{M,q}(X_t^M,t)=\nabla_x\widehat b_{M,q}(X_t^M,t)\,\mathrm dB_t$.
By \eqref{eq:reuse_drift_jacobian} and the It\^o isometry,
\[
 \|\widehat b_{M,q}(X_u^M,u)-\widehat b_{M,q}(X_v^M,v)\|_{L^2}
 \le\sqrt d\,L_\varepsilon\sqrt{u-v},
 \qquad 0\le v\le u\le T_h.
\]
Let $e_k=\|Y_k-X_{t_k}^M\|_{L^2}$.
Subtracting the Euler update from the integral equation and using Minkowski's inequality, we obtain
\[
 e_{k+1}\le(1+L_\varepsilon\Delta t_k)e_k
 +\tfrac23\sqrt d\,L_\varepsilon(\Delta t_k)^{3/2},\qquad k<j.
\]
Since $e_0=0$ and $\sum_{k<j}(\Delta t_k)^{3/2}\le T_h\sqrt h$, the discrete Gronwall inequality yields
\begin{equation}
 e_j\le\tfrac23\sqrt d\,L_\varepsilon T_he^{L_\varepsilon T_h}\sqrt h.
 \label{eq:reuse_prefix_error}
\end{equation}

It remains to control the terminal interval $[T_h,1]$. If $n=1$, then $m_M\equiv Z_1$, and the last update gives $Y_K=Z_1+B_1-B_{t_{K-1}}$, so that $\mathbb E\|Y_K-X_1^M\|^2\le dh$.
Now let $n\ge2$, set $\delta_*=\min_{i\ne l}\|Z_i-Z_l\|>0$, and fix $0<\ell<\delta_*/8$.
If $\|x-Z_i\|\le2\ell$, then $\|Z_l-x\|\ge\frac34\|Z_l-Z_i\|$ and $\|Z_i-x\|\le\frac14\|Z_l-Z_i\|$ for $l\ne i$, so
\[
 \|Z_l-x\|^2-\|Z_i-x\|^2\ge\tfrac12\|Z_l-Z_i\|^2.
\]
Since $m_M-Z_i=\sum_{l\ne i}w_l(Z_l-Z_i)$ and $w_l\le w_l/w_i$, it follows that
\begin{equation}
 \|m_M(x,t)-Z_i\|
 \le\Gamma_M\exp\!\left(-\frac{\delta_*^2}{4(1-t)}\right),\qquad
 \Gamma_M=\max_i\sum_{l\ne i}\frac{c_l}{c_i}\|Z_l-Z_i\|,
 \label{eq:reuse_endpoint_concentration}
\end{equation}
whenever $\|x-Z_i\|\le2\ell$.
By Lemma~\ref{lem:reuse_exact}, $X_1^M=Z_I$ for a random index $I$ with $\mathbb P(I=i)=a_{i,M}$.
Set $\eta_\varepsilon=\Gamma_M\exp[-\delta_*^2/(8\varepsilon)]$, and take $\varepsilon$ small enough that $\eta_\varepsilon\le\ell/4$.
Consider the event
\[
 G_{h,\varepsilon}=\left\{
 \|Y_j-X_{T_h}^M\|\le\ell/4,\quad
 \|X_{T_h}^M-Z_I\|\le\ell/4,\quad
 \sup_{T_h\le u\le1}\|B_u-B_{T_h}\|\le\ell/8
 \right\}.
\]
By \eqref{eq:reuse_time_cutoff} and $1-T_h\le2\varepsilon$, we have $\mathbb E\|X_{T_h}^M-Z_I\|^2\le4R^2\varepsilon^2+2d\varepsilon$.
By Markov's inequality and Doob's maximal inequality, we have
\begin{equation}
 \mathbb P(G_{h,\varepsilon}^c)
 \le\frac{16e_j^2+64R^2\varepsilon^2+544d\varepsilon}{\ell^2}.
 \label{eq:reuse_tail_event}
\end{equation}
We note that this bound does not use independence between $I$ and the subsequent Brownian increments.

Write $\alpha_k=\Delta t_k/(1-t_k)\in(0,1]$.
The remaining Euler updates satisfy
\begin{equation}
 Y_{k+1}-Z_I=(1-\alpha_k)(Y_k-Z_I)
 +\alpha_k\{m_M(Y_k,t_k)-Z_I\}
 +(B_{t_{k+1}}-B_{t_k}),\qquad k\ge j.
 \label{eq:reuse_tail_recursion}
\end{equation}
Iterating \eqref{eq:reuse_tail_recursion} from $j$ to $k$, we find that the noise increments $B_{t_{l+1}}-B_{t_l}$, $j\le l<k$, enter with the coefficients $\prod_{u=l+1}^{k-1}(1-\alpha_u)$, which are deterministic, nondecreasing in $l$, and between zero and one.
By Abel summation, the noise contribution is bounded by $2\sup_{T_h\le u\le1}\|B_u-B_{T_h}\|$.
The drift residuals enter with nonnegative coefficients whose sum is at most one.
Hence, if $Y_j,\ldots,Y_{k-1}$ lie within distance $2\ell$ of $Z_I$, then \eqref{eq:reuse_endpoint_concentration} bounds each drift residual by $\eta_\varepsilon$, and on $G_{h,\varepsilon}$,
\[
 \|Y_k-Z_I\|\le\|Y_j-Z_I\|+\eta_\varepsilon
 +2\sup_{T_h\le u\le1}\|B_u-B_{T_h}\|
 \le\frac\ell2+\frac\ell4+\frac\ell4=\ell.
\]
By induction, on $G_{h,\varepsilon}$ the iterates never leave the $2\ell$-neighborhood of $Z_I$.
Since $\alpha_{K-1}=1$, the final update is $Y_K=m_M(Y_{K-1},t_{K-1})+B_1-B_{t_{K-1}}$, and $\|m_M(Y_{K-1},t_{K-1})-Z_I\|$ is at most $\eta_\varepsilon$ on $G_{h,\varepsilon}$ and at most $2R$ everywhere.
Therefore,
\begin{equation}
 \mathbb E\|Y_K-X_1^M\|^2
 \le2\eta_\varepsilon^2+8R^2\mathbb P(G_{h,\varepsilon}^c)+2dh.
 \label{eq:reuse_terminal_coupling}
\end{equation}

We now conclude. For fixed $\varepsilon$, $e_j\to0$ as $h\downarrow0$ by \eqref{eq:reuse_prefix_error}.
Taking the limit superior as $h\downarrow0$ in \eqref{eq:reuse_terminal_coupling} and using \eqref{eq:reuse_tail_event}, we obtain
\[
 \limsup_{h\downarrow0}\mathbb E\|Y_K-X_1^M\|^2
 \le2\eta_\varepsilon^2+\frac{8R^2(64R^2\varepsilon^2+544d\varepsilon)}{\ell^2}.
\]
Letting $\varepsilon\downarrow0$, the right-hand side tends to zero. This proves the desired result.
\end{proof}

Based on Lemmas~\ref{lem:reuse_exact} and~\ref{lem:reuse_euler}, we prove Theorem~\ref{thm:reuse_posterior_terminal}.

\begin{proof}[Proof of Theorem~\ref{thm:reuse_posterior_terminal}]
Part~(i) follows from Lemmas~\ref{lem:reuse_exact} and~\ref{lem:reuse_euler}. It remains to prove part~(ii).
Write $m_2(\pi)=\int_{\R^d}\|z\|^2\pi(z)\,\mathrm dz$.
We first prove \eqref{eq:reuse_empirical_w2}. By Tonelli's theorem and \eqref{eq:likelihood_expectation}, we have, for bounded measurable $g$,
\[
 \mathbb E_qv_1=\mathcal Z,\qquad
 \mathbb E_q[v_1g(Z_1)]=\mathcal Z\int g(z)\pi(z)\,\mathrm dz,\qquad
 \mathbb E_q[v_1\|Z_1\|^2]=\mathcal Z m_2(\pi),
\]
where the last identity uses nonnegativity.
All these variables are integrable.
Applying the strong law of large numbers simultaneously to a countable convergence-determining class of bounded continuous functions and to $v_1\|Z_1\|^2$, we obtain, almost surely,
\[
 \frac1M\sum_{i=1}^Mv_i\to\mathcal Z>0,\qquad
 \nu_M\Rightarrow\pi,\qquad
 \int\|z\|^2\nu_M(\mathrm dz)\to m_2(\pi).
\]
In particular, $\sum_{i=1}^Mv_i>0$ for all sufficiently large $M$.
Since weak convergence together with convergence of second moments is equivalent to convergence in $W_2$, this proves \eqref{eq:reuse_empirical_w2}.
By the triangle inequality,
\[
 W_2\!\left(\mathcal L(\widetilde X_1^{M,h}\mid\mathcal B_M),\pi\right)
 \le W_2\!\left(\mathcal L(\widetilde X_1^{M,h}\mid\mathcal B_M),\nu_M\right)+W_2(\nu_M,\pi),
\]
and \eqref{eq:reuse_iterated_terminal} follows from \eqref{eq:reuse_euler_terminal} and \eqref{eq:reuse_empirical_w2}.

For the diagonal limit, let $D_{M,K}$ denote the conditional $W_2$ distance in \eqref{eq:reuse_euler_terminal} on the uniform grid with $K$ steps.
On the event $\{\sum_{i=1}^Mv_i>0\}$, $D_{M,K}\to0$ as $K\to\infty$ for each fixed $M$.
By bounded convergence, we can choose deterministic increasing integers $K_M\ge M$ such that
\[
 \mathbb P\!\left(\sum_{i=1}^M v_i>0,\ D_{M,K_M}>1/M\right)\le2^{-M}.
\]
By the Borel--Cantelli lemma and the eventual positivity of $\sum_iv_i$, almost surely $D_{M,K_M}\le1/M$ for all sufficiently large $M$.
Together with \eqref{eq:reuse_empirical_w2} and the triangle inequality, this proves \eqref{eq:reuse_diagonal_terminal}.
\end{proof}

\subsection{Proof of Corollary~\ref{cor:reuse_unconditional_terminal}}
\label{sec:proof_reuse_unconditional}

\begin{proof}
We first prove that $\mathbb EW_2^2(\nu_M,\pi)\to0$. Set $S_M=\sum_{i=1}^Mv_i$, $R_M=\max_{i\le M}\|Z_i\|$, and $J_M=\int\|z\|^2\nu_M(\mathrm dz)$.
Since $v_1>0$ almost surely, $S_M>0$ and $\nu_M$ is defined for every $M$.
We first show that $(J_M)$ is uniformly integrable.
On $G_M=\{S_M\ge M\mathcal Z/2\}$, we have
\[
 J_M\le\frac2{\mathcal Z}\frac1M\sum_{i=1}^M v_i\|Z_i\|^2,
\]
and the sample average on the right converges in $L^1$, so it is uniformly integrable.
On $G_M^c$, we use $J_M\le R_M^2\le\sum_{i=1}^M\|Z_i\|^2$.
Since $v_i\ge0$, we have $S_M\ge\sum_{l\ne i}v_l$, and independence gives
\[
 \mathbb E[R_M^2\mathbf1_{G_M^c}]
 \le M\mathbb E_q\|Z_1\|^2\,
 \mathbb P\!\left(\sum_{i=1}^{M-1}v_i<M\mathcal Z/2\right)
 \le M\mathbb E_q\|Z_1\|^2\,\vartheta^{M-1},\qquad M\ge3,
\]
for some $\vartheta<1$.
Indeed, since $\mathbb E v_1=\mathcal Z>0$, there is $\theta>0$ such that $\vartheta:=e^{3\theta\mathcal Z/4}\mathbb E e^{-\theta v_1}<1$, and the bound follows from the exponential Markov inequality and $M/2\le3(M-1)/4$ for $M\ge3$.
Hence the contribution of $G_M^c$ tends to zero in $L^1$, and $(J_M)$ is uniformly integrable.
Since $W_2^2(\nu_M,\pi)\le2J_M+2m_2(\pi)$, \eqref{eq:reuse_empirical_w2} and uniform integrability yield $\mathbb EW_2^2(\nu_M,\pi)\to0$.

Next, for fixed $M$, coupling the last Euler update with an independent draw from $\nu_M$, conditionally on $\mathcal B_M$, gives
\[
 D_{M,h}^2:=W_2^2\!\left(\mathcal L(\widetilde X_1^{M,h}\mid\mathcal B_M),\nu_M\right)
 \le4R_M^2+dh.
\]
Since $\mathbb ER_M^2\le M\mathbb E_q\|Z_1\|^2<\infty$, \eqref{eq:reuse_euler_terminal} and dominated convergence give $\mathbb ED_{M,h}^2\to0$ as $h\downarrow0$.
Mixing the conditional couplings over $\mathcal B_M$, we obtain
\[
 W_2\!\left(\mathcal L(\widetilde X_1^{M,h}),\pi\right)
 \le(\mathbb ED_{M,h}^2)^{1/2}
 +(\mathbb EW_2^2(\nu_M,\pi))^{1/2},
\]
which proves the iterated limit.
For the diagonal limit, we choose the deterministic $K_M$ larger if necessary so that, in addition to the probability bound in the proof of Theorem~\ref{thm:reuse_posterior_terminal}, $\mathbb ED_{M,1/K_M}^2\le2^{-M}$.
This proves the desired result.
\end{proof}

\subsection{Proof of Theorem~\ref{thm:qbank_pointwise}}
\label{sec:proof_pointwise}

\begin{proof}
We suppress the dependence on the fixed $(x,t)$ and define
\begin{equation}
 \chi_i^q
 =\frac{\Omega_q(Z_i,\xi_i;x,t)}{C(x,t)}
 \left\{\frac{Z_i-x}{1-t}-b(x,t)\right\},
 \qquad
 \widehat C_{M,q}=\frac1M\sum_{i=1}^M
 \Omega_q(Z_i,\xi_i;x,t).
 \label{eq:qbank_influence}
\end{equation}
By \eqref{eq:qbank_change_measure} and \eqref{eq:qbank_exact_drift}, $\mathbb E_q\chi_i^q=0$, and by \eqref{eq:qbank_centered_moment}, $\mathbb E_q\|\chi_i^q\|^2<\infty$.
Direct subtraction of the two ratios gives
\begin{equation}
 \widehat b_{M,q}-b
 =
 \frac{C(x,t)}{\widehat C_{M,q}}
 \frac1M\sum_{i=1}^M\chi_i^q.
 \label{eq:qbank_centered_ratio}
\end{equation}
By the strong law of large numbers, $\widehat C_{M,q}\to C(x,t)>0$ and $M^{-1}\sum_i\chi_i^q\to0$ almost surely.
Hence $\widehat b_{M,q}$ is well defined for all sufficiently large $M$, and \eqref{eq:qbank_slln} follows from \eqref{eq:qbank_centered_ratio}.
By the multivariate central limit theorem,
\begin{equation}
 M^{-1/2}\sum_{i=1}^M \chi_i^q
 \Rightarrow
 \mathcal N\!\left(0,\mathbb E_q[\chi_1^q(\chi_1^q)^\top]\right).
 \label{eq:qbank_influence_clt}
\end{equation}
Since $\chi_i^q=\Omega_q(Z_i,\xi_i;x,t)(Z_i-\bar z)/[(1-t)C(x,t)]$, Slutsky's theorem applied to \eqref{eq:qbank_centered_ratio} gives \eqref{eq:qbank_clt} with covariance \eqref{eq:qbank_asymcov}.
This proves the desired result.
\end{proof}

\subsection{Variance and effective sample size}
\label{sec:proposal_variance_details}

By \eqref{eq:qbank_asymcov} and the definition of $\Omega_q$,
\[
 \mathbb E_q\!\left[\Omega_q^2\|Z-\bar z\|^2\right]
 =\int q(z)\frac{\varphi_{1-t}(z-x)^2}{q(z)^2}
 \mathbb E_{\xi\mid z}\!\left[W(z,\xi)^2\|z-\bar z\|^2\right]\mathrm dz
 =\mathcal V_{x,t}(q),
\]
which gives \eqref{eq:qbank_trace}.
For any proposal $q$ satisfying the support condition, the Cauchy--Schwarz inequality gives
\begin{equation}
 \left(\int\sqrt{\mathcal A_{x,t}(z)}\,\mathrm dz\right)^2
 \le
 \left(\int\frac{\mathcal A_{x,t}(z)}{q(z)}\,\mathrm dz\right)
 \left(\int q(z)\,\mathrm dz\right)
 =
 \mathcal V_{x,t}(q),
 \label{eq:qbank_optimal_cs}
\end{equation}
with equality for $q=q^*_{x,t}$ in \eqref{eq:qbank_optimal_proposal} whenever $q^*_{x,t}$ satisfies the support condition.
Finally, $q_3\ge0.05q_1$ gives $\mathcal A_{x,t}/q_3\le20\mathcal A_{x,t}/q_1$, and integrating proves \eqref{eq:qbank_defensive_bound}.

\begin{remark}[Local ESS and terminal importance weights]
\label{rem:reuse_ess}
In the bridge representation of Lemma~\ref{lem:reuse_exact}, Bayes' formula and the Gaussian density identity in its proof give
\begin{equation}
 \mathbb P(I=i\mid X_t^M=x,\mathcal B_M)
 =\frac{c_i\varphi_{1-t}(Z_i-x)}{\sum_j c_j\varphi_{1-t}(Z_j-x)}.
 \label{eq:reuse_endpoint_conditional}
\end{equation}
Thus \eqref{eq:qbank_ess} measures the concentration of the conditional endpoint probabilities.
In contrast, the effective sample size of the terminal importance weights is
\[
 \operatorname{ESS}_{\mathrm{term}}
 =\left(\sum_i a_{i,M}^2\right)^{-1}
 =\operatorname{ESS}(0,0).
\]
If the nodes with positive weights are distinct, then $X_t^M\to Z_I$ almost surely and, for $j\ne I$,
\[
 \frac{c_j\varphi_{1-t}(Z_j-X_t^M)}{c_I\varphi_{1-t}(Z_I-X_t^M)}
 =\frac{c_j}{c_I}\exp\!\left\{-\frac{\|Z_j-X_t^M\|^2-\|Z_I-X_t^M\|^2}{2(1-t)}\right\}
 \longrightarrow0.
\]
Consequently, $\operatorname{ESS}(X_t^M,t)\to1$ almost surely; for repeated nodes, the same conclusion holds after combining the weights at each distinct node.
A low local ESS near time one therefore does not by itself indicate posterior error, a lack of independent outputs, or numerical instability, since it also occurs under exact sampling of $\nu_M$.
\end{remark}

\bibliographystyle{IEEEtran}
\bibliography{references}

\end{document}